# Analysis of evolutionary origins of genomic loci harboring 59,732 candidate human-specific regulatory sequences identifies genetic divergence patterns during evolution of Great Apes

Gennadi V. Glinsky[1]

[1] Institute of Engineering in Medicine

University of California, San Diego

9500 Gilman Dr. MC 0435

La Jolla, CA 92093-0435, USA

Correspondence: gglinskii@ucsd.edu

Web: http://iem.ucsd.edu/people/profiles/guennadi-v-glinskii.html

**Running title:** Multispecies mosaicism of human-specific regulatory sequences

**Abstract**

Contemporary knowledge of the universe of genomic regions harboring various types of candidate human-specific regulatory sequences (HSRS) has been markedly expanded in recent years. To infer the evolutionary origins of genomic loci harboring HSRS, analyses of sequence conservations patterns of 59,732 HSRS-harboring regions in genomes of Modern Humans, Chimpanzee, Bonobo, Gorilla, Orangutan, Gibbon, and Rhesus have been performed. Two major evolutionary pathways have been identified comprising thousands of genomic sequences that were either inherited from extinct common ancestors (ECAs) or created de novo in human genomes after the split of human and chimpanzee lineages. Present analyses revealed thousands of HSRS that appear inherited from ECAs yet bypassed genomes of our closest evolutionary relatives, Chimpanzee and Bonobo, presumably due to the incomplete lineage sorting and/or species-specific loss or regulatory DNA. The bypassing pattern is particularly prominent for HSRS putatively associated with development and functions of human brain. Common genomic loci that are likely contributed to speciation during evolution of Great Apes have been identified comprising of 248 insertions sites of African Great Ape-specific retrovirus PtERV1 (45.9%; p = 1.03E-44) intersecting genomic regions harboring 442 HSRS, which are enriched for HSRS that have been associated with human-specific (HS) changes of gene expression in cerebral organoid models of brain development. Among non-human primates (NHP), most significant fractions of candidate HSRS associated with HS gene expression changes in both excitatory neurons (347 loci; 67%) and radial glia (683 loci; 72%) are highly conserved in the Gorilla genome. Present analyses revealed that Modern Humans acquired and maintained unique combinations of regulatory sequences that are highly conserved in distinct species of six NHP separated by 30 million years of evolution. Concurrently, this unique-to-human mosaic of genomic regulatory sequences inherited from ECAs was supplemented with 12,486 created de novo HSRS. Evidence of multispecies evolutionary origins of HSRS support the model of complex continuous speciation process during evolution of Great Apes that is not likely to occur as an instantaneous event.

## Introduction

Recent advances enabled by the analyses of individual genomes of Great Apes using high-resolution sequencing technologies and methodologically diverse comparative analyses of human and non-human primates' reference genomes significantly enhanced our understanding of human-specific structural genomic variations of potential regulatory and functional significance (Locke et al., 2005; Chimpanzee Sequencing and Analysis Consortium, 2005; McLean et al., 2011; Prüfer et al., 2012; Shulha et al., 2012; Konopka et al., 2012; Scally et al., 2012; Capra et al., 2013; Marchetto et al., 2013; Marnetto et al., 2014; Prescott et al., 2015; Gittelman et al. 2015; Glinsky et al., 2015-2018; Dong et al., 2016; Sousa et al., 2017; Dennis et al., 2017; Kronenberg et al., 2018; Guffani et al., 2018). Collectively, these studies markedly expanded the observable universe of candidate human-specific regulatory sequences (HSRS), which currently comprises nearly sixty thousand genomic loci aligned to the most recent release of the human reference genome (Table 1; Supplemental Tables S1-S3). This remarkable progress highlights a multitude of significant contemporary challenges, the centerpiece of which is a need to compile a comprehensive catalog of HSRS in order to identify the high-priority panel of genetic targets for stringent functional validation experiments. The selected high-priority genetic panel of human phenotypic divergence would represent the elite set of HSRS, which will be chosen based on the expectation of high-likelihood of biologically-significant species-specific effects on phenotypes that would be revealed during in-depth structural-functional explorations of their impact on development of human-specific traits.

One of essential steps toward addressing this problem is to gain insights into evolutionary origins of genomic regions harboring HSRS. In this contribution, an up-to-date catalog of 59,732 candidates human-specific regulatory loci has been assembled and their conservation patterns in genomes of five non-human Great Apes (Chimpanzee, Bonobo, Gorilla, Orangutan, and Gibbon; Tables 1-6) have been analyzed. To identify the putative hot spots of the Great Apes' genetic divergence, systematic comparisons of genomic coordinates of HSRS and unique to African Great Apes insertions of the PtERV1 retrovirus-derived sequences were carried-out by performing comprehensive genome-wide proximity placement analyses. Diverse patterns of sequence conservation of different classes of HSRS were observed, reflecting quantitatively distinct profiles of inheritance from extinct common ancestors (ECAs) of the human lineage and each of the five species of non-human Great Apes. One of the prevalent modes of sequence conservation is represented by the bypassing pattern of evolutionary inheritance, which is exemplified by thousands of HSRS that appear inherited from ECAs yet bypassed genomes of our closest evolutionary relatives, Chimpanzee and Bonobo. The bypassing pattern of evolutionary inheritance seems particularly prominent for candidate HSRS putatively associated with development and functions of human brain. Reported herein the putative hot spots of genetic divergence of Modern Humans represent the elite set of high-priority candidates for in-depth structural-functional characterization of their contribution to evolution of human-specific traits.

## Results

### Insertion sites of the African Great Ape-specific retrovirus PtERV1 and significant fractions of distinct classes HSRS share common genomic coordinates

Structurally distinct mutations within genomic regions harboring HSRS that independently emerged on the Modern Humans lineage and distinct species of non-human Great Apes are of particular interest because they might indicate the functional divergence between species of these independently-targeted regulatory regions. In this context, it was of interest to determine whether genomic regions harboring HSRS intersect genomic coordinates of insertion sites of the African Great Ape-specific retrovirus PtERV1. Significantly, no PtERV1 insertions were detected in genomes of Modern Humans and Orangutan yet the PtERV1 retrovirus appears integrated at 540 loci in genomes of Gorilla, Chimpanzee, and Bonobo during millions years of evolution (Kronenberg et al., 2018). Analysis of evolutionary patterns of insertions of African ape-specific retrovirus PtERV1 revealed that, without exception, all distinct classes of HSRS analyzed in this study intersect within 10 Kb windows orthologous genomic regions targeted by PtERV1 in genomes of Gorilla, Chimpanzee, and Bonobo, albeit with different degrees of frequencies and significance (Tables 1-2; Figure 1). Interestingly, genomic coordinates of PtERV1 insertions in the Gorilla genome appear to overlap more frequently genomic regions harboring HSRS (Table 2; Figure 1).

Genome-wide analysis of gene expression changes in human-chimpanzee cerebral organoids revealed that human-specific duplications, in contrast to other types of human-specific structural variations, are associated with up-regulated genes in human radial glial and excitatory neurons (Kronenberg et al., 2018). These observations are highly consistent with previous studies demonstrating that defined human-specific segmental duplications of *SRGAP2* and *ARHGAP11B* genes drive phenotypic differences in cortical development between humans and chimpanzee (Dennis et al., 2012; Charrier et al., 2012; Florio et al., 2015). Proximity placement enrichment analysis of 7,897 duplication regions in human genome and PtERV1 insertion sites (Table 2a) identified 71 PtERV1 loci intersecting 87 duplication regions and revealed significantly more frequent co-localization of Chimpanzee-specific PtERV1 insertions compared to Gorilla-specific insertions (24.2% versus 10.7%, respectively; p = 0.0076; 2-tailed Fisher's exact test). Overall, these analyses identified 248 PtERV1 loci (45.9%; p = 1.03E-44; hypergeometric distribution test) intersecting human genomic regions harboring 442 candidate HSRS (Tables 1-2), which are significantly enriched (p = 0.0018) for regions of fixed human-specific mutations that have been associated with human-specific changes of gene expression in cerebral organoids' models of brain development (Kronenberg et al., 2018). This set of genomic regions with overlapping coordinates of PtERV1 integration sites and loci harboring human-specific mutations of potential functional significance may represent an attractive functional validation panel of elite candidate regulatory sequences likely contributing to phenotypic divergence of Modern Humans and our closest evolutionary relatives.

**Mosaicism of evolutionary origins of genomic loci harboring various classes of human-specific mutations**

Recent experiments identified 24,151 genomic regions harboring various classes of human-specific mutations identified based on the comparative analyses of genomes of Modern Humans and non-human Great Apes (Kronenber et al., 2018). It was of interest to analyze the sequence conservation patterns of these regions in genomes of six non-human primates (NHP), including five non-human Great Apes (Chimpanzee, Bonobo, Gorilla, Orangutan, and Gibbon) and Rhesus Macaque (Table 3). In these analyses, genomic sequences that manifested at least 95% of sequence conservations during the direct and reciprocal conversions from/to reference genomes of Modern Humans (hg38) and corresponding NHP species were considered highly-conserved. Within the context of definition of evolutionary origins of genomic regions harboring human-specific mutations, one of the main motivations was the inference that this analytical effort would identify highly-conserved DNA sequences that were inherited by the Modern Humans' lineage from ECAs.

Consistent with a model of the significant contribution of the ECA's inheritance, a majority (66%-88%) of 19,221 genomic loci harboring various classes of human-specific mutations appears highly conserved in genomes of Great Apes and Rhesus (Table 3). In contrast, only 13% of 4,910 human-specific short tandem repeats (STR) expansions' regions are conserved. Consistent with the predominantly primate's origins of regulatory regions harboring human-specific mutations, less than 5% of analyzed sequences appear highly conserved in the mouse genome (Table 3). Two classes of HSRS were mapped to large fractions of DNA sequences highly-conserved in genomes of various species of NHP: DNA loci harboring 49.6 to 80.6% of fixed human-specific deletions and 59.2 to 79.6% of human-specific short tandem repeats (STR) contractions were identified as highly-conserved genomic regions in genomes of different species of non-human Great Apes and Rhesus (Table 3).

Overall, the fractions of highly-conserved sequences harboring human-specific mutations and assigned to different NHP's species seem to reflect the consensus evolutionary order of the NHP genomes' similarity to the genome of Modern Humans. Most notable exceptions from this pattern were identified during the analyses of human-specific mutations associated with human-specific gene expression changes in excitatory neurons and radial glia (Tables 3 & 4). Among NHP species, a markedly prominent majority of genomic regions harboring human-specific mutations associated with human-specific gene expression changes during brain development in both excitatory neurons and radial glia is highly conserved only in genomes of our three closest evolutionary relatives: 50.9% and 50.3% in Chimpanzee; 39.9% and 47.5% in Bonobo; 67.1% and 72.1% in Gorilla (for excitatory neurons and radial glia, respectively). In contrast, only ~ 5% of genomic regions harboring human-specific mutations associated with human-specific gene expression changes detected in cerebral organoids are highly-conserved in genomes of Orangutan, Gibbon, and Rhesus. Among non-human Great Apes, most significant fractions of candidate HSRS associated with human-specific gene expression changes in both excitatory neurons (347 loci; 67%) and radial glia (683 loci; 72%) are highly conserved in the Gorilla genome. For highly-conserved regions harboring HSRS associated with human-specific gene expression changes in excitatory neurons, differences in conservation profiles between genomes of Gorilla, Chimpanzee, and Bonobo were highly significant as defined by the two-tailed Fisher's exact test (p = 1.434E-07; p = 1.574E-18; p = 0.000463). For highly-conserved regions harboring HSRS associated with human-specific gene expression changes in radial glia, differences in conservation profiles between genomes of Gorilla and Chimpanzee as well as Gorilla and Bonobo were highly significant (p = 1.623E-22 and p = 7.369E-28, respectively). In contrast, conservation profiles of HSRS associated with human-specific gene expression changes in radial glia were similar in genomes of Chimpanzee and Bonobo (p = 0.250; Table 3). These observations suggest that the majority of highly-conserved genomic regions harboring candidate HSRS associated with human-specific differences of gene expression in both excitatory neurons and radial glia was inherited from ECAs of Modern Humans and Gorilla. This conclusion remains valid when the analyses were performed considering either only loci remapped to/from NHP's genomes to identical hg38 genomic coordinates (Table 4; Figures 2-4; Supplemental Table S4) or only genomic loci uniquely mapped to genomes of only single species of non-human Great Apes (Figure 3-4). Notably, differences in conservation profiles of genomic loci harboring HSRS associated with human-specific gene expression changes in radial glia appear particularly prominent (Table 4; Figures 2-4; Supplemental Table S4). Consistent with the ECA's inheritance model, from 85.3% to 95.1% of HSRS-harboring regions that are highly-conserved in the genomes of Chimpanzee and Bonobo are highly-conserved in the Gorilla genome as well. In contrast, only from 59.0% to 59.3% and from 62.7% to 65.1% of HSRS-harboring regions that are highly-conserved in the Gorilla genome remain highly-conserved in the genomes of Bonobo and Chimpanzee, respectively.

**Mosaicism of evolutionary origins of candidate human-specific regulatory loci defined based on the mapping failure to both Chimpanzee and Bonobo reference genomes**

One of the approaches to the identification of candidate HSRS is based on their absence in the genomes of our closest evolutionary relatives, Chimpanzee and Bonobo. Present analyses demonstrate that large fractions of genomic regions harboring candidate HSRS are highly-conserved in genomes of our more distant evolutionary relatives, Gorilla, Orangutan, Gibbon, and Rhesus (Tables 3-4). These observations suggest that candidate human-specific regulatory loci which were defined based on the mapping failures to both Chimpanzee and Bonobo genomes may originate on DNA sequences highly-conserved in genomes of other

NHP species. To test the validity of this hypothesis, 16,730 genomic loci harboring distinct classes of HSRS were identified that failed to convert to genomes of both Chimpanzee and Bonobo using 10% sequence identity threshold (Table 5). Then highly-conserved sequences in genomes of Rhesus, Gibbon, Orangutan, and Gorilla were identified and tabulated for each category of HSRS. Consistent with the bypassing patterns of evolutionary inheritance, thousands of distinct classes of candidate HSRS that failed to map to genomes of both Chimpanzee and Bonobo are highly conserved in genomes of Gorilla, Orangutan, Gibbon, and Rhesus (Table 5). Most prominently, significant fractions of retrotransposon-derived loci that are transcriptionally-active in human dorsolateral prefrontal cortex and absent in genomes of both Chimpanzee and Bonobo are highly conserved in genomes of Gorilla, Orangutan, Gibbon, and Rhesus (1,688; 1,371; 1,148; and 1,045 loci, respectively). It has been observed that in all instances the Gorilla genome had the largest numbers of shared with Modern Humans highly-conserved sequences that failed to map to genomes of both Chimpanzee and Bonobo (Table 5).

These observations indicate that a more stringent approach for definition of candidate HSRS which are likely to have been created de novo in the genome of Modern Humans would be to require the conversion failures of a regulatory DNA sequence to all six NHP's genomes, namely genomes of Chimpanzee, Bonobo, Gorilla, Orangutan, Gibbon, and Rhesus. Using this strategy, 12,486 candidate HSRS have been identified (Table 6), indicating that 24.8% of all analyzed in this contribution HSRS-harboring genomic loci could be classified as created de novo candidate HSRS.

**Enrichment within human-specific regulatory pathways of genes comprising expression signatures of human-specific neurodevelopmental and pluripotency transcriptional networks**

It is possible that distinct types of candidate HSRS are not just unrelated elements of a random population of DNA sequences, but they might represent a coherent collection of regulatory DNA sequences assembled during evolution to facilitate execution of human-specific functions. If this hypothesis is correct, then HSRS may represent the key components of human-specific genomic regulatory pathways governing human-specific gene expression patterns observed in various phenotypic contexts associated with human-specific traits. Therefore, the validity of this hypothesis could be tested using strictly-defined by comparisons with non-human

Great Apes human-specific gene expression signatures associated with development of human brain in the cerebral organoid model (Kronenberg et al., 2018) and induced pluripotency phenotypes of human versus NHP cells (Marchetto et al., 2013). Kronenberg et al. (2018) identified several hundred genes that manifest human-specific expression changes in Modern Humans versus Chimpanzee cerebral organoids' model of brain development. Significant fractions of these human-specific neurodevelopmental networks operating in both excitatory neurons and radial glia appear associated with human-specific structural variants, specifically, with human-specific insertions and deletions (Table 7). Even larger fractions of genes comprising human-specific neurodevelopmental networks have been identified as components of the gene expression signature (GES) of the MLME cells in human preimplantation embryo: common gene sets represent 262 genes (68.2%; p = 1.59E-93) and 481 genes (72.0%; p = 3.95E-187) for human-specific GES of excitatory neurons and radial glia, respectively (Table 7).

It has been reported that the creation of the MLME cells in human preimplantation embryos is associated with increased expression of primate-specific retrotransposon-derived regulatory long non-coding RNAs termed human pluripotency-associated transcripts, HPATs (Glinsky et al., 2018). Most recently, the expansive networks of primate-specific and human-specific retrotransposons transcriptionally active in human dorsolateral prefrontal cortex (DLPFC) and associated coding genes have been identified (Guffani et al., 2018). Thus, it was of interest to determine what fractions of genes comprising human-specific neurodevelopmental gene expression networks in excitatory neurons and radial glia may overlap with coding genes coupled with active transcription of transposable elements in human DLPFC. Remarkably, it has been observed that common gene sets represent a vast majority of genes comprising human-specific neurodevelopmental networks: they comprise 322 genes (83.9%; p = 5.57E-84) and 561 genes (84%; p = 3.08E-146) for human-specific GES of excitatory neurons and radial glia, respectively (Table 7). Interestingly, both *SRGAP2C* and *ARHGAP11B* genes driving divergent cortical development between humans and chimpanzee (Dennis et al., 2012; Charrier et al., 2012; Florio et al., 2015) harbor transposable elements transcriptionally active in human DLPFC (Guffani et al., 2018).

Marchetto et al. (2013) identified human-specific gene expression signature distinguishing induced pluripotent stem cells (iPSC) engineered from cells of Modern Humans and NHP species. To infer the putative regulatory association patterns of genes distinguishing human iPSC versus NHP iPSC and human-specific genomic regulatory pathways, overlap enrichment analyses of corresponding gene sets have been performed. It has been observed, that the association patterns of human-specific gene expression signatures of brain development and pluripotency phenotypes with compendiums of genes likely governed by human-specific regulatory pathways appear strikingly similar (Tables 7 and 8). Overall, 88% of genes comprising human-specific expression signature of the induced pluripotency phenotype represent genes implicated in putative regulatory associations with human-specific genomic pathways. These observations support the hypothesis that human-specific gene expression signatures of brain development and pluripotency phenotypes are associated with a collection of HSRS assembled during evolution into human-specific genomic regulatory pathways to govern transcriptional networks in human cells.

**Discussion**

An impressive contemporary collection of nearly sixty thousand candidate HSRS assembled by the collective decades-long effort of many laboratories (see Introduction) lends further credence to the idea that unique to human phenotypes might result from human-specific changes to genomic regulatory sequences (King and Wilson, 1975). This study identifies multiple high-priority candidate HSRS for in depth structural-functional validation analyses, among which most prominent candidate HSRS appears associated with human-specific gene expression changes in excitatory neurons and radial glia as well as in human induced pluripotent stem cells. This high-priority set of elite genetic targets include candidate HSRS putatively regulating expression of *SERINC5*, *APOBEC3B*, and *PIWIL2* genes, high expression of which in human cells is likely to confer increased resistance to the retroviral infection and propagation of retrotransposons (Marchetto et al., 2013; Usami et al., 2015; Rosa et al., 2015). It is tempting to speculate that these changes may have been significant genetic contributors conferring the selective fitness advantage to human lineage during primate evolution.

High-confidence human-specific mutations leading to emergence of candidate HSRS should be considered as rare genomic events that unlikely to occur more than once during evolution at the same genomic locations. Therefore, observations that large fractions of genomic regions harboring distinct classes of HSRS represent DNA sequences highly-conserved in genomes of distinct NHP species should be interpreted as strong circumstantial evidence consistent with their putative regulatory functions. Collectively, the evidence presented in this contribution revealed a complex unique-to-human mosaic of regulatory DNA sequences inherited from ECAs following separation events from multiple distinct NHP species and reflecting the striking ancestral polymorphism of Modern Humans. One of the novel mechanisms that may have contributed to divergence of genomic regulatory networks of Modern Humans and non-human Great Apes is illustrated by observations that the insertions of the African Great Ape-specific retrovirus PtERV1 and distinct classes of HSRS have common genomic coordinates within orthologous genomic regions of Gorilla, Chimpanzee, Bonobo, and Modern Humans. Overall, these common patterns of species-specific mutations within overlapping genomic regions were observed for 248 PtERV1 insertions and 442 HSRS, including 21 HSRS associated with genes differentially expressed in human versus chimpanzee cerebral organoid models of brain development.

Observations reported herein support the hypothesis that the speciation process during evolution of Great Apes is not likely to occur as an instantaneous event (Patterson et al., 2006): for example, human and chimpanzee lineages could have exchanged genes following the iterative sequences of the initial lineage divergence, separation, and gathering together prior to the permanent segregation of two species. Since human, chimpanzee, and gorilla lineages may have diverged during the relatively short evolutionary time, this model might reflect the extended complex speciation process of these three closely-related Great Apes, possibly involving co-evolution of their ECAs. Incomplete lineage sorting events were intrinsic components of the genomic divergence and are likely played an important role in the lineage segregation. In agreement with this hypothesis, comparative analyses of multiple alignments of sequences of human, chimpanzee, gorilla, and orangutan genomes have demonstrated that a considerable fraction of genes in the human genome is more similar to the gorilla genome than to the chimpanzee genome (Patterson et al., 2006; Chen and Li, 2001; Yang, 2002; O'hUigin et al., 2002; Wall, 2003; Hobolth et al. 2007). Genomic regions harboring HSRS appear to follow the similar evolutionary trajectory. Observed examples of the bypassing pattern of the evolutionary inheritance highlight HSRS supporting the inference of alternative genealogies (human being most closely related to NHP other than Chimpanzee) are most likely reflect the incomplete lineage sorting events.

Incomplete lineage sorting has been consistently observed in the multiple alignments of the genomes for human, chimpanzee, gorilla, and orangutan where differences in models of gene genealogies and species phylogeny were documented for up to 36% of the human autosomal genome (Chen and Li, 2001; Yang, 2002; Wall, 2003; Patterson et al. 2006; Hobolth et al., 2007; 2011; Kronenberg et al., 2018). Similar changes could result from species-specific losses of conserved ancestral loci of regulatory DNA, a mechanism that contributed to evolution of human-specific traits (McLean et al., 2011).

**Conclusion**

Observations reported in this contribution support the conclusion that Modern Humans captured unique combinations of human-specific regulatory loci, divergent subsets of which were created within genomic regions highly conserved in distinct species of six NHP separated by 30 million years of evolution. Concurrently, this unique-to-human mosaic of genomic regulatory pathways built on DNA sequences inherited from ECAs was supplemented with 12,486 created de novo HSRS. Collectively, present findings suggest that incremental genomic divergence of the human lineage has been continued throughout the primate's evolution concurrently with the emergence and segregation of other non-human Great Ape species. This complex continuous process of genomic divergence was gradually driving speciation of *H. sapiens*, in part, by capturing and retaining the unique mosaic of genomic signatures of ECAs.

**Methods**

**Data source**

*Candidate human-specific regulatory sequences and African Apes-specific retroviral insertions*

A total of 51,835 candidate HSRS and all currently known 504 insertion sites of the African Apes-specific PtERV1 retrovirus were analyzed in this study, detailed descriptions of which and corresponding references of primary original contributions are reported in the Tables 1-8 and Supplemental Tables S1-S5.

*Additional Data Sources and Analytical Protocols*

Solely publicly available datasets and resources were used in this contribution as well as methodological approaches and a computational pipeline validated for discovery of primate-specific gene and human-specific regulatory loci (Tay et al., 2009; Kent, 2002; Schwartz et al., 2003; Capra et al., 2013; Marnetto et al., 2014; Glinsky, 2015-2018; Guffani et al., 2018). The analysis is based on the University of California Santa Cruz (UCSC) LiftOver conversion of the coordinates of human blocks to corresponding non-human genomes using chain files of pre-computed whole-genome BLASTZ alignments with a minMatch of 0.95 and other search parameters in default setting (http://genome.ucsc.edu/cgi-bin/hgLiftOver). Extraction of BLASTZ alignments by the LiftOver algorithm for a human query generates a LiftOver output "Deleted in new", which indicates that a human sequence does not intersect with any chains in a given non-human genome. This indicates the absence of the query sequence in the subject genome and was used to infer the presence or absence of the human sequence in the non-human reference genome. Human-specific regulatory sequences were manually curated to validate their identities and genomic features using a BLAST algorithm and the latest releases of the corresponding reference genome databases for time periods between April, 2013 and September, 2018.

The significance of the differences in the expected and observed numbers of events was calculated using two-tailed Fisher's exact test. Additional placement enrichment tests were performed for individual classes of HSRS taking into account the size in bp of corresponding genomic regions. Datasets of NANOG-, POU5F1-, and CTCF-binding sites and human-specific TFBS in hESCs as well as all other classes of HSRS were reported previously (Kunarso et al., 2010; McLean et al., 2011; Prüfer et al., 2012; Shulha et al., 2012; Konopka et al., 2012; Scally et al., 2012; Capra et al., 2013; Marchetto et al., 2013; Marnetto et al., 2014; Prescott et al., 2015; Gittelman et al. 2015; Glinsky et al., 2015-2018; Dong et al., 2016; Sousa et al., 2017; Dennis et al., 2017; Kronenberg et al., 2018; Guffani et al., 2018) and are publicly available.

**Data analysis**

**Categories of DNA sequence conservation**

Identification of highly-conserved in primates (pan-primate), primate-specific, and human-specific sequences was performed as previously described (Glinsky, 2015-2018). In brief, all categories were defined by direct and reciprocal mapping using liftOver (see above). Specifically:

- Highly conserved in primates' sequences: DNA sequences that have at least 95% of bases remapped during conversion from/to human (Homo sapiens, hg38), chimp (Pan troglodytes, v5), and bonobo (Pan paniscus, v2). Similarly, highly-conserved sequences were defined for hg38 and genomes of Gorilla, Orangutan, Gibbon, and Rhesus.
- Primate-specific: DNA sequences that failed to map to the mouse genome (mm10).
- Human-specific: DNA sequences that failed to map at least 10% of bases from human to both chimpanzee and bonobo. All candidate HSRS identified based on the sequence alignments failures to genomes of both chimpanzee and bonobo were subjected to more stringent additional analyses requiring the mapping failures to genomes of Gorilla, Orangutan, Gibbon, and Rhesus. These loci were considered created de novo human-specific regulatory sequences (HSRS).

Additional comparisons were performed using the same methodology and exactly as stated in the manuscript text and described in details below.

**Genome-wide proximity placement analysis**

Genome-wide Proximity Placement Analysis (GPPA) of distinct genomic features co-localizing with HSRS was carried out as described previously (Glinsky, 2015-2018). Briefly, as typical example of the analytical protocol, we examined the significance of overlaps between hESC active enhances and hsTFBS by first identifying all hsTFBS that overlap with any of the genomic regions tested in the ChIP-STARR-seq dataset (Barakat etl, 2018; Glinsky et al., 2018). We then calculated the relative frequency of active enhancers overlapping with hsTFBS. To assess the significance of the observed overlap of genomic coordinates, we compared the values recorded for hsTFBS with the expected frequency of active and non-active enhancers that overlap with all TFBS for NANOG (15%) and OCT4 (25%) as previously determined (Barakat et al 2018). Our analyses demonstrate that more than 95% of hsTFBS co-localized with sequences in the tested regions of the hESC genome. Analyses of conservation patterns of 11,866 human-specific insertions have been performed using eleven different window sizes (Supplemental Table S6) centered at the insertion sites previously reported by Kronenberg et al. (2018). Numbers of records that successfully completed direct and reciprocal conversions from/to hg38 and genomes of non-human species (six non-human primates and mouse) using sequence identity threshold 95% are reported in the Supplemental Table S6.

**Evolutionary origin and functional enrichment analyses**

Evolutionary origins of HSRS were inferred from the results of the conservation patterns of 59,732 candidate human-specific regulatory DNA sequences based on the hg38 release of the human reference genome and latest available releases of genomes of six non-human primates, namely Chimpanzee, Bonobo, Gorilla, Orangutan, Gibbon, and Rhesus. The conservation analyses was carried-out using the LiftOver algorithm and Multiz Alignments of 20 mammals (17 primates) of the UCSC Genome Browser (Kent et al., 2002) on Human Dec. 2013 Assembly (GRCh38/hg38) (http://genome.ucsc.edu/cgi-bin/hgTracks?db=hg38&position=chr1%3A90820922-90821071&hgsid=441235989_eelAivpkubSY2AxzLhSXKL5ut7TN ).

All DNA sequences were converted to most recent releases of the corresponding reference genome databases and were utilized consistently throughout the study to ensure the use of the most precise, accurate, and reproducible genomic DNA sequences available to date. A candidate HSRS was considered conserved if it could be aligned from/to hg38 reference genome and either one or both *Chimpanzee* or *Bonobo* genomes using defined sequence conservation thresholds of the LiftOver algorithm MinMatch function and direct and reciprocal conversions protocols. Similarly, the conservation patterns were evaluated for genomes of other NHP. LiftOver conversion of the coordinates of human blocks to non-human genomes using chain files of pre-computed whole-genome BLASTZ alignments with a specified MinMatch levels and other search parameters in default setting (http://genome.ucsc.edu/cgi-bin/hgLiftOver). Several thresholds of the LiftOver algorithm MinMatch function (minimum ratio of bases that must remap) were utilized to assess the sequences conservation and identify candidate human-specific (MinMatch of 0.1; 0.95; 0.99; and 1.00) and conserved in nonhuman primates (MinMatch of 0.95 and 1.00) regulatory sequences as previously described (Glinsky, 2015-2018; Guffani et al., 2018). The Net alignments provided by the UCSC Genome Browser were utilized to compare the sequences in the human genome (hg38) with the mouse (mm10), *Chimpanzee* (PanTro5), and latest available releases of *Bonobo*, Gorilla, *Orangutan*, *Gibbon*, and *Rhesus* genomes. A given regulatory

DNA segment was defined as the highly conserved regulatory sequence when both direct and reciprocal conversions between humans' and nonhuman primates' genomes were observed using the MinMatch sequence alignment threshold of 0.95 requiring that 95% of bases must remap during the alignments of the corresponding sequences. A given regulatory DNA segment was defined as the created de novo candidate human-specific regulatory sequence when sequence alignments failed to both *Chimpanzee* and *Bonobo* genomes using the specified MinMatch sequence alignment thresholds. More stringently, these requirements were extended to include genomes of Gorilla, Orangutan, Gibbon, and Rhesus. Analyses of conservation patterns of 11,866 human-specific insertions have been performed using eleven different window sizes (Supplemental Table S6) centered at the insertion sites previously reported by Kronenberg et al. (2018). Numbers of records that successfully completed direct and reciprocal conversions from/to hg38 and genomes of non-human species (six non-human primates and mouse) using sequence identity threshold 95% are reported in the Supplemental Table S6.

The Enrichr API (January 2018 version) (Chen et al., 2013) was used to test genes linked to HSRS of interest for significant enrichment in numerous functional categories. To comply with the web interface, we considered the 1000 genes closest to the tested peaks for enrichments. In all plots, we report the "combined score" calculated by Enrichr, which is a product of the significance estimate and the magnitude of enrichment (combined score $c = log(p) * z$, where $p$ is the Fisher's exact test p-value and $z$ is the z-score deviation from the expected rank). Additional functional enrichment analyses were performed with GREAT (McLean et al., 2010).

*Statistical Analyses of the Publicly Available Datasets*

All statistical analyses of the publicly available genomic datasets, including error rate estimates, background and technical noise measurements and filtering, feature peak calling, feature selection, assignments of genomic coordinates to the corresponding builds of the reference human genome, and data visualization, were performed exactly as reported in the original publications and associated references linked to the corresponding data visualization tracks (http://genome.ucsc.edu/). Any modifications or new elements of statistical analyses are described in the corresponding sections of the Results. Statistical significance of the Pearson correlation coefficients was determined using GraphPad Prism version 6.00 software. The significance of the differences in the numbers of events between the groups was calculated using two-sided Fisher's exact and Chi-square test, and the significance of the overlap between the events was determined using the hypergeometric distribution test (Tavazoie et al., 1999).

## Supplemental Information

Supplemental information includes Supplemental Tables S1-S5.

## Author Contributions

This is a single author contribution. All elements of this work, including the conception of ideas, formulation, and development of concepts, execution of experiments, analysis of data, and writing of the paper, were performed by the author.

## Acknowledgements

This work was made possible by the open public access policies of major grant funding agencies and international genomic databases and the willingness of many investigators worldwide to share their primary research data. I would like to thank my anonymous colleagues for their valuable critical contributions during the peer review process of this work.

**Figure legends**

**Figure 1.** Mosaicism of evolutionary origins of 7,897 duplicated regions in the hg38 release of human reference genome defined by whole-genome shotgun sequence detection (WSSD).

A. A consensus model of the lineage speciation during the evolution of Great Apes. The arrows depict the hypothetical flow of genomic information inherited from extinct common ancestors (ECAs) and acquired through species-specific gain and losses. Evolutionary origins via ECA's inheritance pathways of highly-conserved sequences in genomes of Modern Humans and non-human species of Great Apes are postulated.
B. Mosaicism of evolutionary origins of 7,897 duplicated regions in the hg38 release of human reference genome defined by WSSD (Kronenberg et al., 2018). Numbers of highly-conserved regions that successfully completed direct and reciprocal conversion tests are reported for each non-human species (NHS).
C. Species-specific mosaicism of evolutionary origins of duplicated regions in the hg38 release of human reference genome defined by WSSD. Only highly-conserved sequences unique for each genome of NHS are reported.
D. Linear regression analysis of duplication regions in the hg38 release of human reference genome defined by WSSD that are highly conserved in genomes of NHS. Numbers of highly-conserved sequences and numbers of lost ancestral loci are shown for each genome of NHS.

**Figure 2.** Mosaicism of evolutionary origins of candidate human-specific and primate-specific regulatory loci defined by sequence conservation analyses.

A. Evolutionary patterns of inheritance, gains, and losses of 540 insertions of Africa Great Apes-specific retrovirus PtERV1.
B. Evolutionary patterns of inheritance, gains, and losses of 947 human-specific regulatory regions associated with human-specific changes of gene expression in radial glia. Only records identically remapped loci in NHP during direct and reciprocal conversions from/to hg38 are reported.

C. Evolutionary patterns of inheritance, gains, and losses of 517 human-specific regulatory regions associated with human-specific changes of gene expression in excitatory neurons. Only records identically remapped loci in NHP during direct and reciprocal conversions from/to hg38 are reported.

D. Evolutionary patterns of inheritance, gains, and losses of 4645 human-specific transposable elements (TE) loci transcriptionally active in human dorsolateral prefrontal cortex (DLPFC). Common ancestor heritage number of 2361 loci takes into account 1,045 loci highly conserved in Rhesus; 4645 loci failed conversion to PanTro5 & PanPan1 (sequence identity threshold of 10%); 18 loci completed direct & reciprocal conversions (sequence identity threshold of 95%) from/to hg38 & PanPan2; 4612 failed conversions to PanTro5; PanPna1; PanPan2 (sequence identity threshold of 10%).

**Figure 3.** Species-specific mosaicism of evolutionary origins of human-specific regulatory sequences associated with human-specific changes of gene expression in excitatory neurons (A, B) and radial glia (C, D).

A. Highly-conserved sequences reciprocally mapped as identical loci to the 282 regions harboring human-specific mutations associated with human-specific gene expression changes in excitatory neurons.

B. Highly-conserved species-specific sequences reciprocally mapped as identical loci to the genomic regions harboring human-specific mutations associated with human-specific gene expression changes in excitatory neurons.

C. Highly-conserved sequences reciprocally mapped as identical loci to the 525 regions harboring human-specific mutations associated with human-specific gene expression changes in radial glia.

D. Highly-conserved species-specific sequences reciprocally mapped as identical loci to the genomic regions harboring human-specific mutations associated with human-specific gene expression changes in radial glia.

**Figure 4.** Species-specific mosaicism of evolutionary origins of human-specific regulatory sequences encoded by transcriptionally-active transposable elements (TE) in human DPLFC.

A. Numbers of sequences highly-conserved in NHP genomes reciprocally mapped to human-specific DLPFC-expressed transposable elements.

B. Highly-conserved species-specific sequences reciprocally mapped from NHP genomes to human-specific DLPFC-expressed transposable elements.

Figure 1.

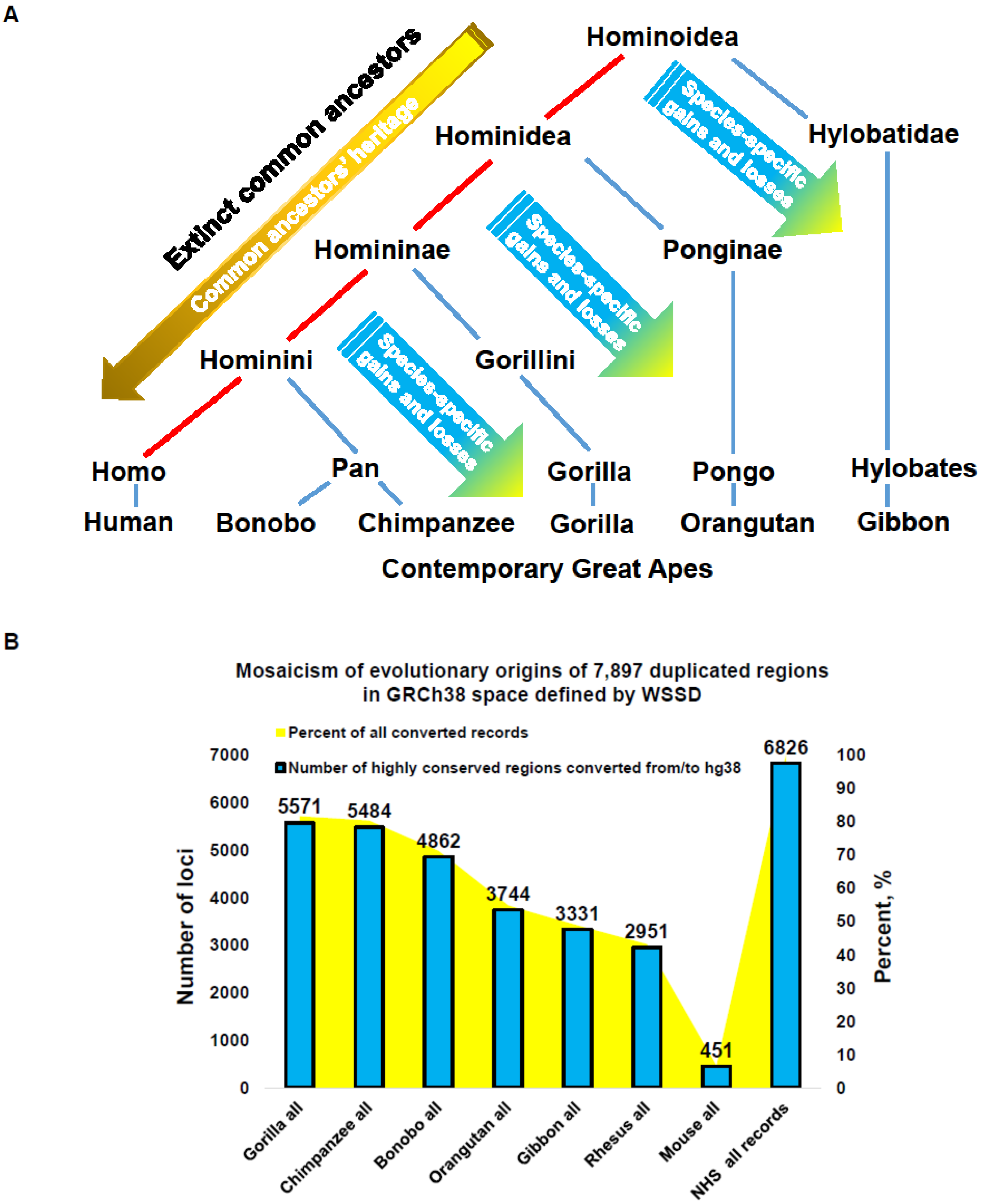
A
Extinct common ancestors
Common ancestors' heritage
Hominoidea
Hominidea
Hylobatidae
Homininae
Ponginae
Hominini
Gorillini
Species-specific gains and losses
Species-specific gains and losses
Species-specific gains and losses
Homo
Pan
Gorilla
Pongo
Hylobates
Human
Bonobo
Chimpanzee
Gorilla
Orangutan
Gibbon
Contemporary Great Apes
B
Mosaicism of evolutionary origins of 7,897 duplicated regions
in GRCh38 space defined by WSSD
Percent of all converted records
Number of highly conserved regions converted from/to hg38
Number of loci
Percent, %
5571
5484
4862
3744
3331
2951
451
6826
Gorilla all
Chimpanzee all
Bonobo all
Orangutan all
Gibbon all
Rhesus all
Mouse all
NHS all records

C

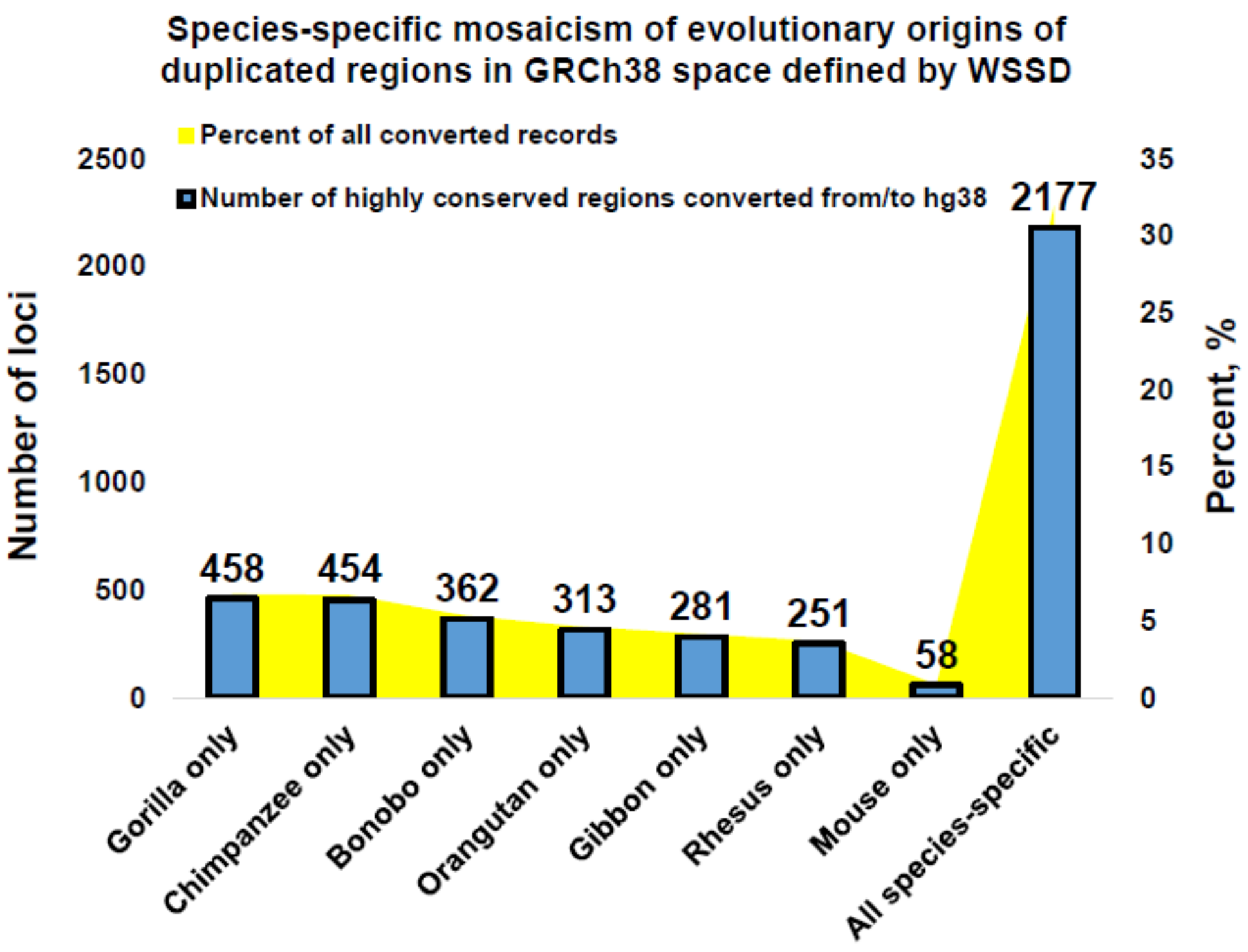
Species-specific mosaicism of evolutionary origins of duplicated regions in GRCh38 space defined by WSSD
Percent of all converted records
Number of highly conserved regions converted from/to hg38
Number of loci
Percent, %
2500
2000
1500
1000
500
0
35
30
25
20
15
10
5
0
458
454
362
313
281
251
58
2177
Gorilla only
Chimpanzee only
Bonobo only
Orangutan only
Gibbon only
Rhesus only
Mouse only
All species-specific

D

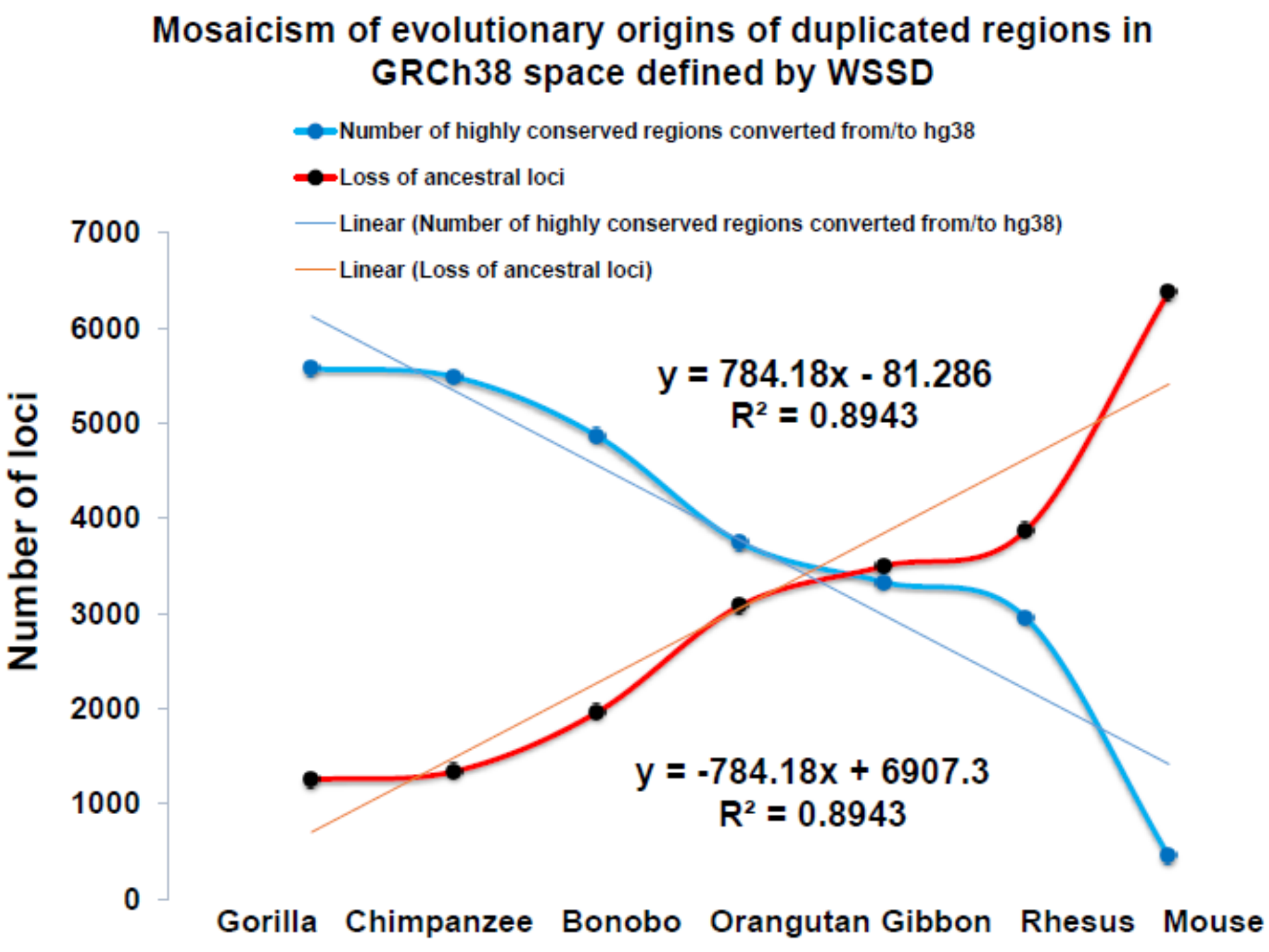
Mosaicism of evolutionary origins of duplicated regions in GRCh38 space defined by WSSD
Number of highly conserved regions converted from/to hg38
Loss of ancestral loci
Linear (Number of highly conserved regions converted from/to hg38)
Linear (Loss of ancestral loci)
Number of loci
7000
6000
5000
4000
3000
2000
1000
0
y = 784.18x - 81.286
R² = 0.8943
y = -784.18x + 6907.3
R² = 0.8943
Gorilla
Chimpanzee
Bonobo
Orangutan
Gibbon
Rhesus
Mouse

**Figure 2.**

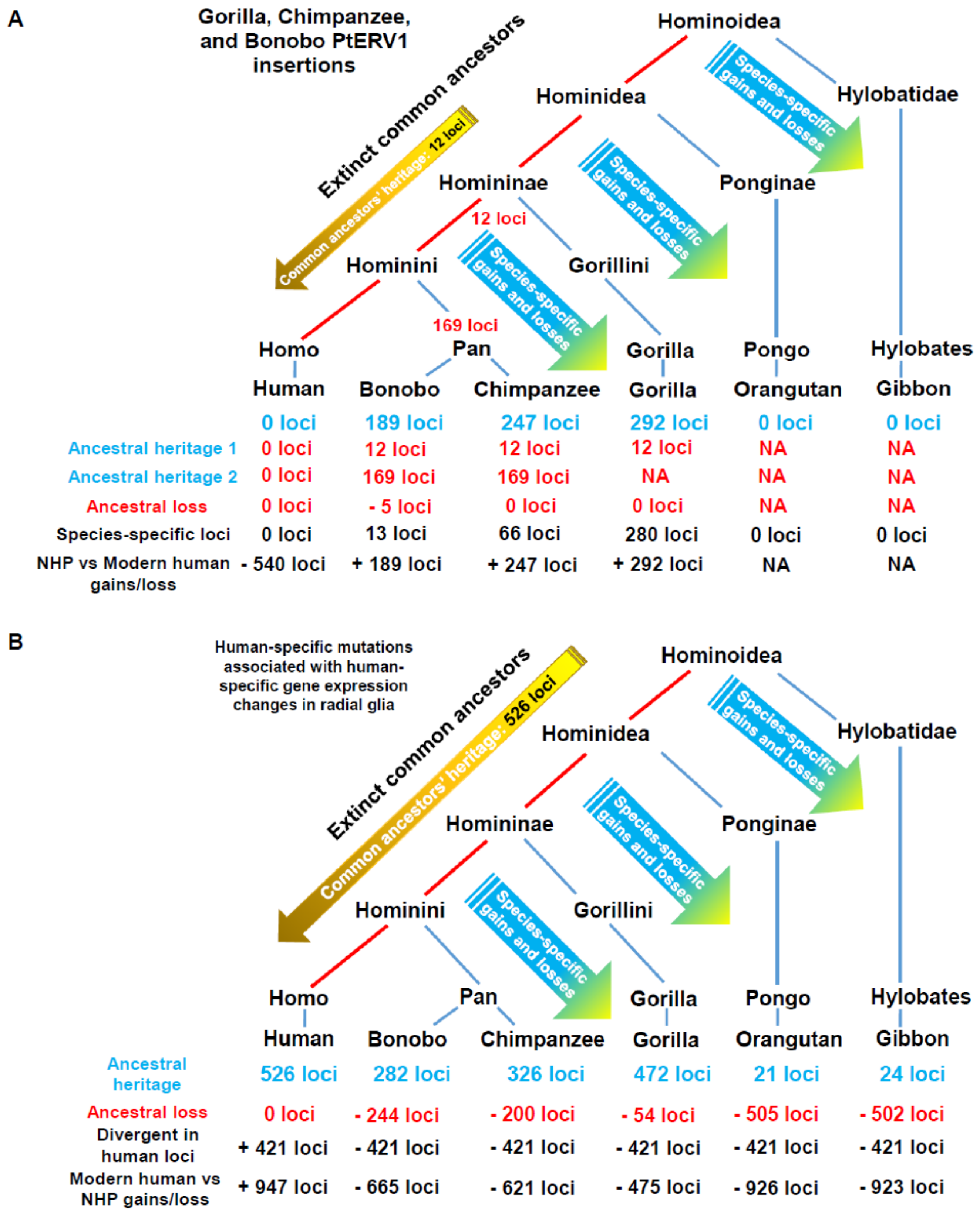
A
Gorilla, Chimpanzee, and Bonobo PtERV1 insertions
Extinct common ancestors
Common ancestors' heritage: 12 loci
Hominoidea
Hominidea
Hylobatidae
Homininae
Ponginae
Hominini
Gorillini
12 loci
169 loci
Species-specific gains and losses
Homo
Pan
Gorilla
Pongo
Hylobates
Human
Bonobo
Chimpanzee
Gorilla
Orangutan
Gibbon
0 loci | 189 loci | 247 loci | 292 loci | 0 loci | 0 loci
Ancestral heritage 1 | 0 loci | 12 loci | 12 loci | 12 loci | NA | NA
Ancestral heritage 2 | 0 loci | 169 loci | 169 loci | NA | NA | NA
Ancestral loss | 0 loci | - 5 loci | 0 loci | 0 loci | NA | NA
Species-specific loci | 0 loci | 13 loci | 66 loci | 280 loci | 0 loci | 0 loci
NHP vs Modern human gains/loss | - 540 loci | + 189 loci | + 247 loci | + 292 loci | NA | NA
B
Human-specific mutations associated with human-specific gene expression changes in radial glia
Extinct common ancestors
Common ancestors' heritage: 526 loci
Hominoidea
Hominidea
Hylobatidae
Homininae
Ponginae
Hominini
Gorillini
Species-specific gains and losses
Homo
Pan
Gorilla
Pongo
Hylobates
Human
Bonobo
Chimpanzee
Gorilla
Orangutan
Gibbon
Ancestral heritage | 526 loci | 282 loci | 326 loci | 472 loci | 21 loci | 24 loci
Ancestral loss | 0 loci | - 244 loci | - 200 loci | - 54 loci | - 505 loci | - 502 loci
Divergent in human loci | + 421 loci | - 421 loci | - 421 loci | - 421 loci | - 421 loci | - 421 loci
Modern human vs NHP gains/loss | + 947 loci | - 665 loci | - 621 loci | - 475 loci | - 926 loci | - 923 loci

C

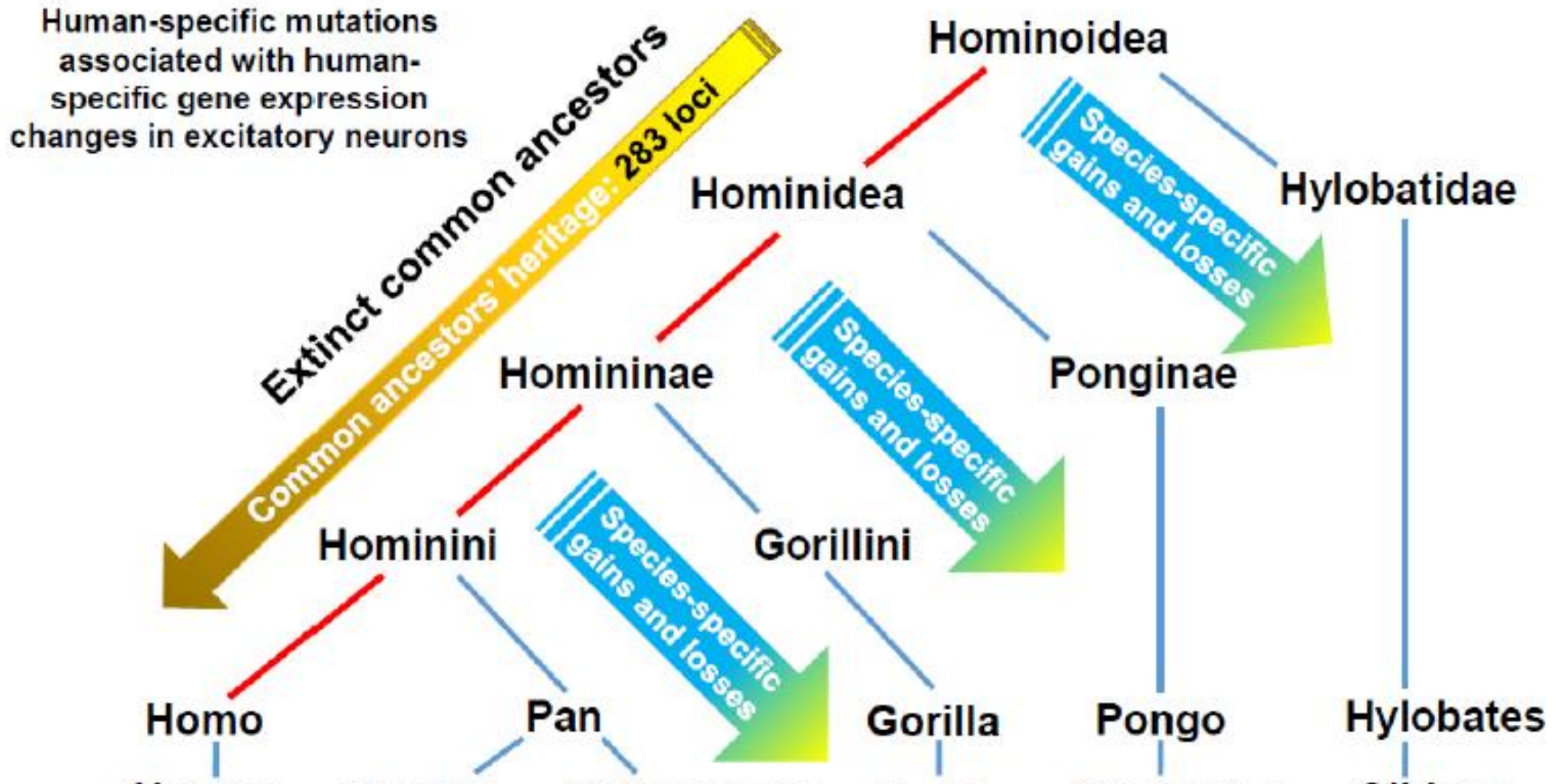

| | Human | Bonobo | Chimpanzee | Gorilla | Orangutan | Gibbon |
|---|---|---|---|---|---|---|
| Ancestral heritage | 283 loci | 143 loci | 183 loci | 249 loci | 9 loci | 16 loci |
| Ancestral loss | 0 loci | - 140 loci | - 100 loci | - 34 loci | - 274 loci | - 267 loci |
| Divergent in human loci | + 234 loci | - 234 loci | - 234 loci | - 234 loci | - 234 loci | - 234 loci |
| Modern human vs NHP gains/loss | + 517 loci | - 374 loci | - 334 loci | - 268 loci | - 508 loci | - 501 loci |

D

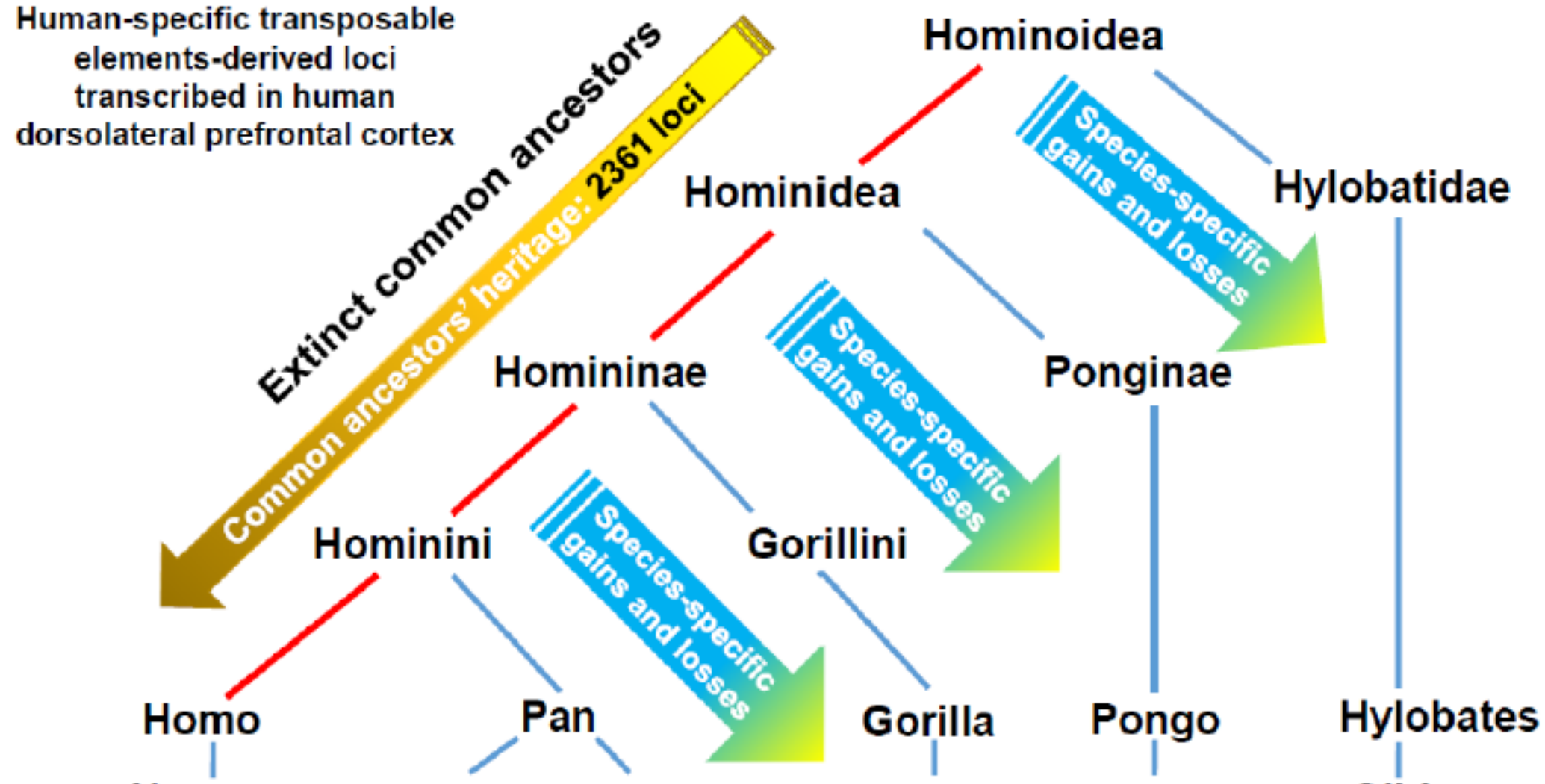

| | Human | Bonobo | Chimpanzee | Gorilla | Orangutan | Gibbon |
|---|---|---|---|---|---|---|
| Ancestral heritage | 2361 loci | 18 loci | 0 loci | 1688 loci | 1371 loci | 1148 loci |
| Ancestral loss | 0 loci | - 2343 loci | - 2361 loci | - 673 loci | - 990 loci | - 1213 loci |
| Human-specific loci | + 2284 loci | - 2284 loci | - 2284 loci | - 2284 loci | - 2284 loci | - 2284 loci |
| Modern human vs NHP gains/loss | + 4645 loci | - 4627 loci | - 4645 loci | - 2957 loci | - 3274 loci | - 3497 loci |

**Figure 3.**

**A**

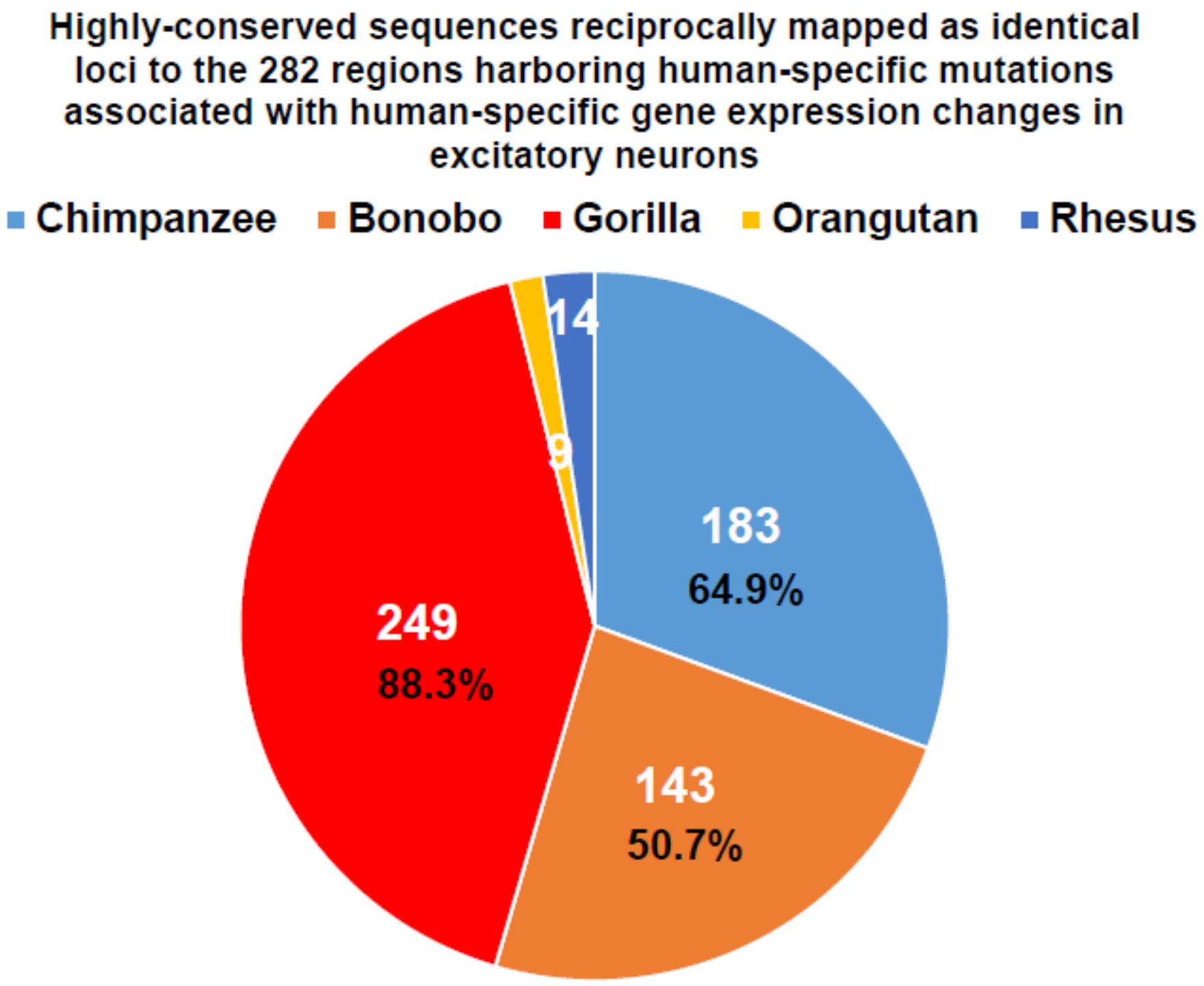

Highly-conserved sequences reciprocally mapped as identical loci to the 282 regions harboring human-specific mutations associated with human-specific gene expression changes in excitatory neurons
Chimpanzee
Bonobo
Gorilla
Orangutan
Rhesus
14
9
183
64.9%
143
50.7%
249
88.3%

**B**

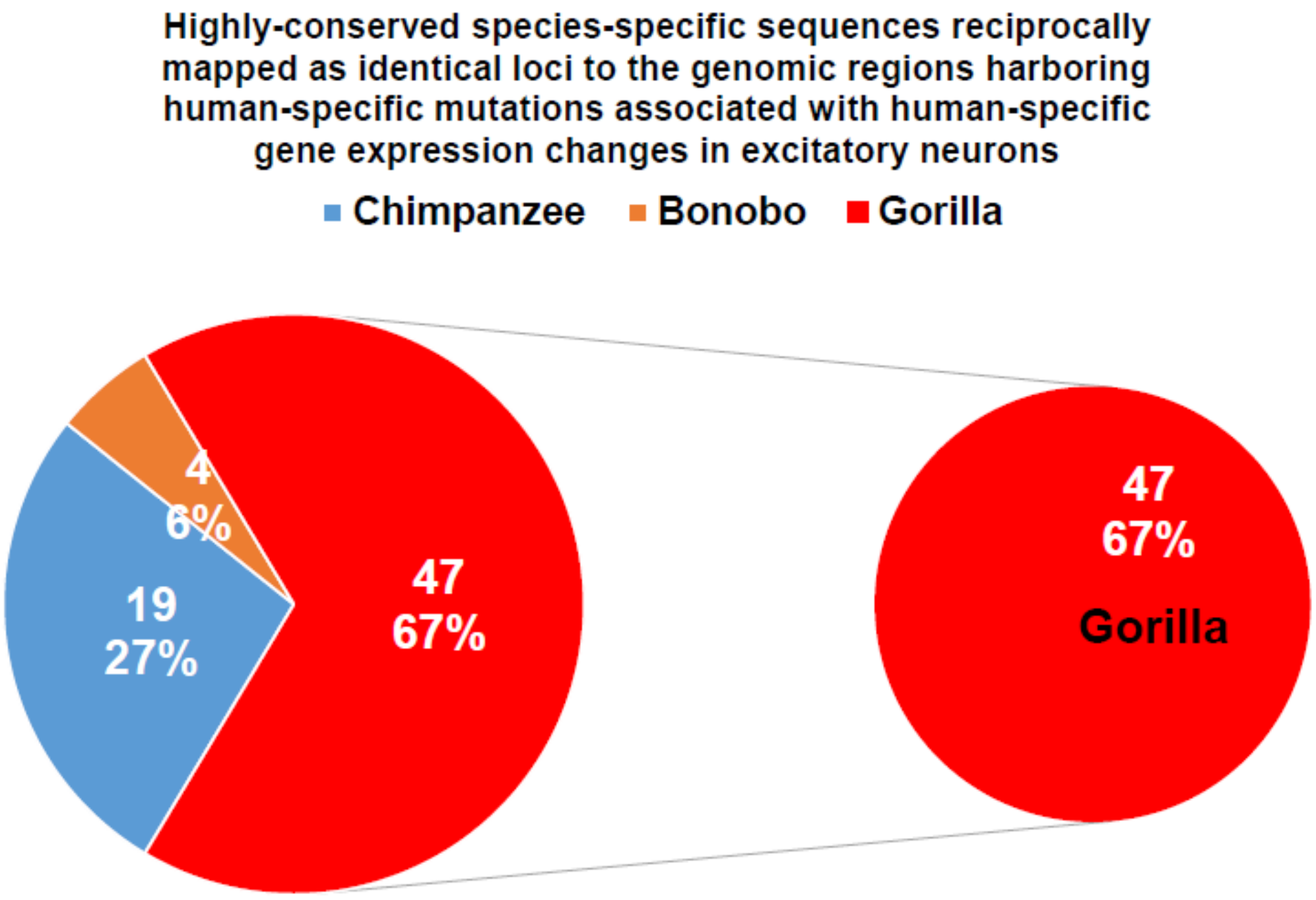

Highly-conserved species-specific sequences reciprocally mapped as identical loci to the genomic regions harboring human-specific mutations associated with human-specific gene expression changes in excitatory neurons
Chimpanzee
Bonobo
Gorilla
4
6%
19
27%
47
67%
47
67%
Gorilla

C

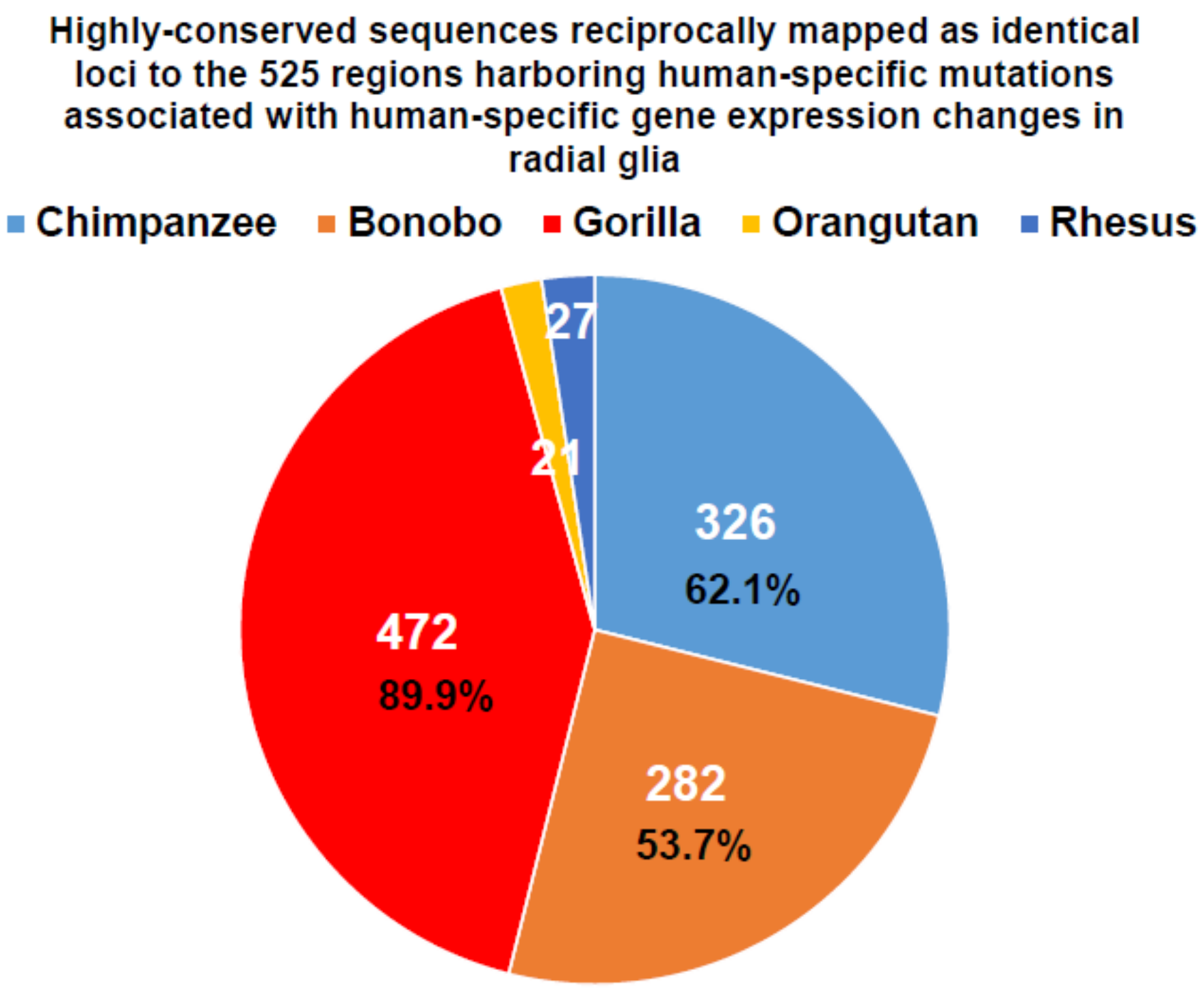
Highly-conserved sequences reciprocally mapped as identical loci to the 525 regions harboring human-specific mutations associated with human-specific gene expression changes in radial glia
Chimpanzee
Bonobo
Gorilla
Orangutan
Rhesus
27
21
326
62.1%
472
89.9%
282
53.7%

D

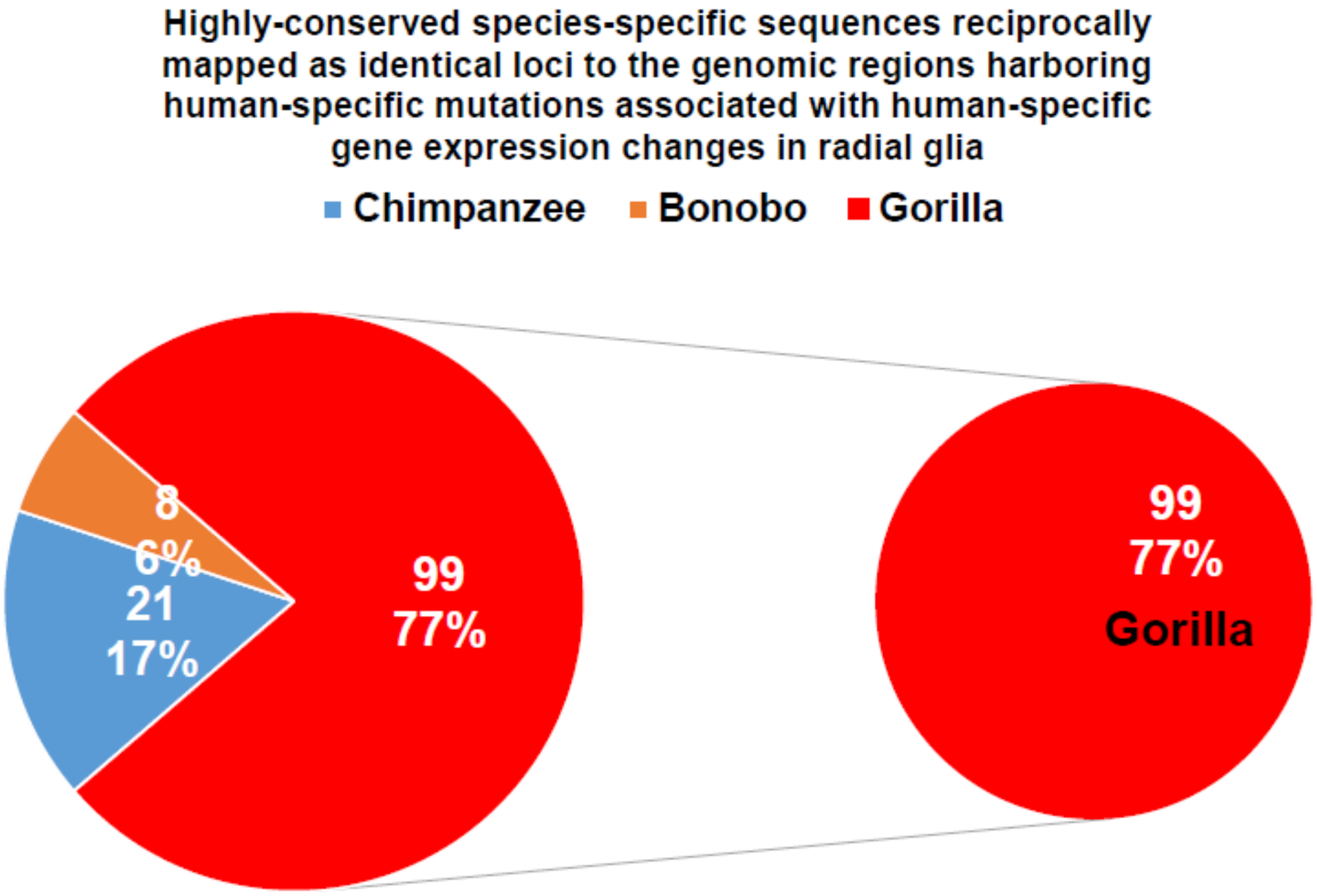
Highly-conserved species-specific sequences reciprocally mapped as identical loci to the genomic regions harboring human-specific mutations associated with human-specific gene expression changes in radial glia
Chimpanzee
Bonobo
Gorilla
8
6%
21
17%
99
77%
99
77%
Gorilla

**Figure 4.**

**A**

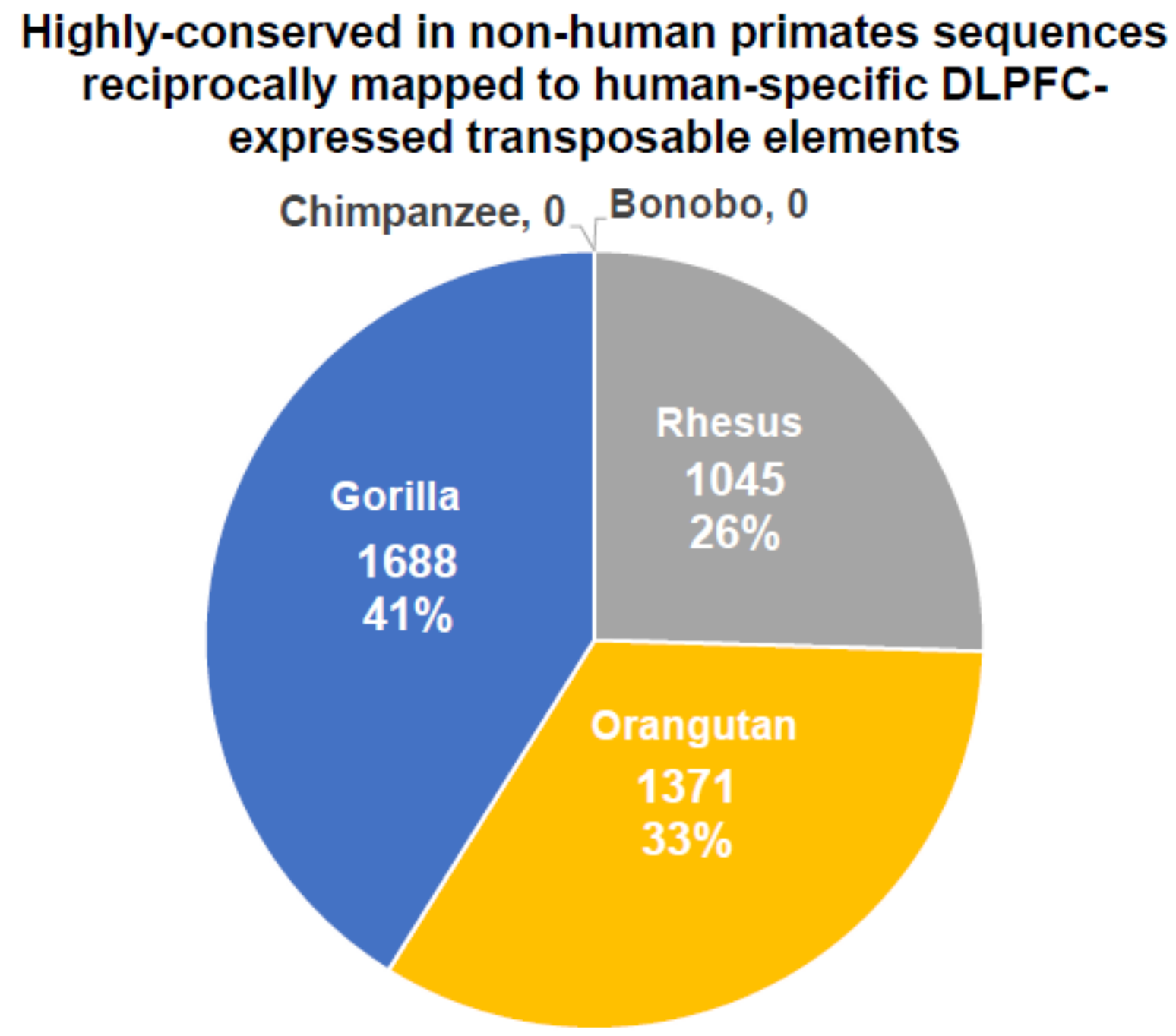

DLPFC, dorsolateral prefrontal cortex

**B**

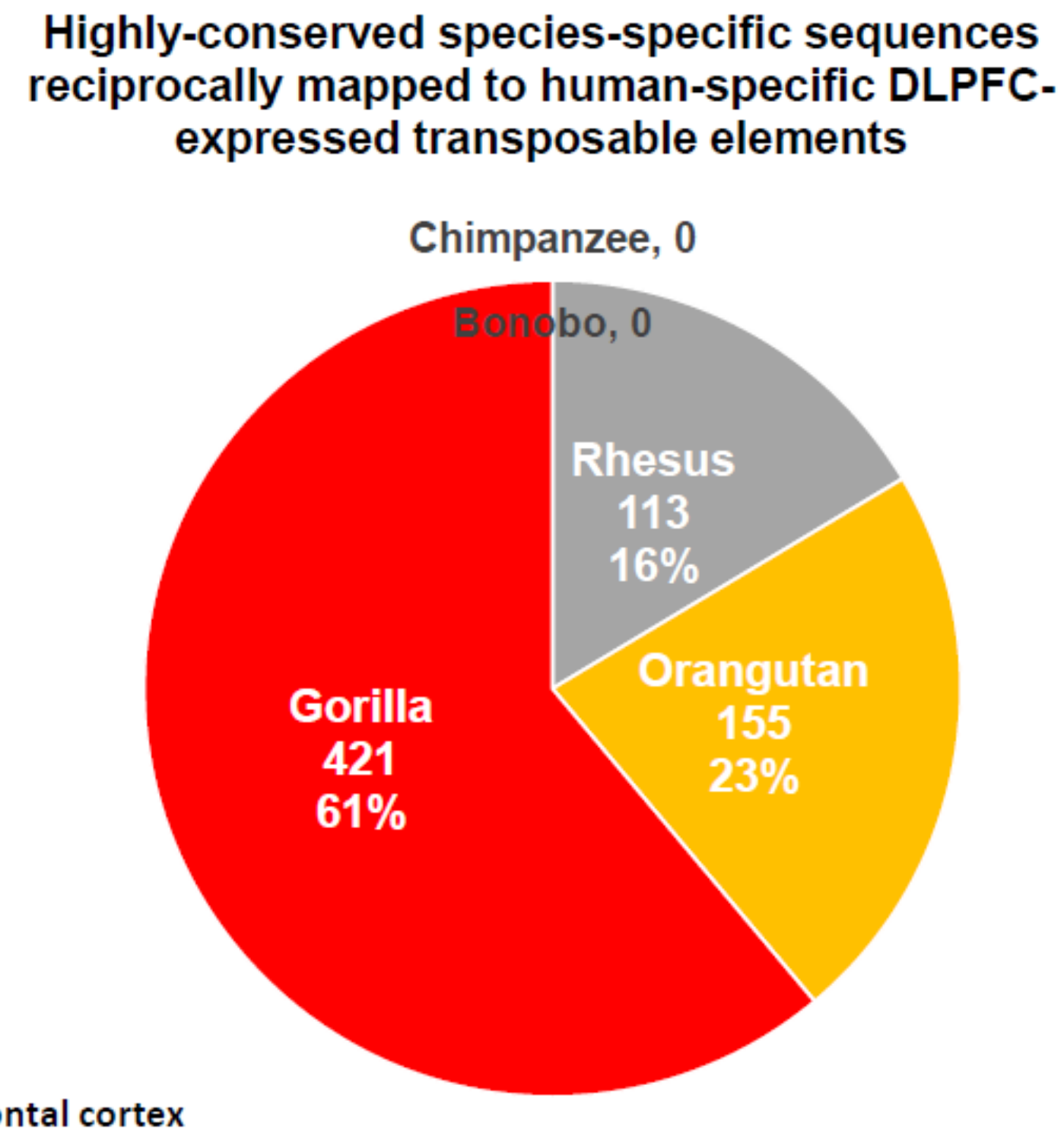

DLPFC, dorsolateral prefrontal cortex

**Table 1.** Distribution profiles of human genomic regions harboring 442 loci of distinct classes of candidate HSRS co-localizing with insertion sites of 248 PtERV1 loci in Chimpanzee, Gorilla, and Bonobo genomes.

| Human-specific regulatory sequences (HSRS) | Number of loci | Intersecting PtERV1 loci | Intersecting human-specific loci | Percent | P value** |
|---|---|---|---|---|---|
| **Human-specific STR contractions*** | 1464 | 4 | 4 | 0.27 | |
| **Human-specific STR expansions** | 4910 | 25 | 25 | 0.51 | 0.374 |
| **Fixed human-specific deletions** | 5891 | 38 | 37 | 0.63 | 0.118 |
| **Fixed human-specific insertions** | 11886 | 66 | 66 | 0.56 | 0.182 |
| **All regions of human-specific mutations** | 24151 | 133 | 132 | 0.55 | 0.195 |
| **Fixed human-specific regulatory regions (FHSRR)** | 4249 | 15 | 31 | 0.73 | 0.053 |
| **Accelerated evolution-DHS (ace-DHS)** | 3538 | 23 | 25 | 0.71 | 0.068 |
| **Human accelerated regions (HARs)** | 2741 | 12 | 11 | 0.40 | 0.597 |
| **Chimp-biased developmental enhancers** | 999 | 5 | 5 | 0.50 | 0.499 |
| **Human segmental duplications** | 218 | 2 | 2 | 0.92 | 0.176 |
| **Duplicated regions in GRCh38 space defined by WSSD** | 7897 | 71 | 87 | 1.10 | 0.001 |
| **Human-biased developmental enhancers** | 996 | 9 | 10 | 1.00 | 0.026 |
| **DHS Fixed human-specific regulatory regions (DHS-FHSRR)*** | 2116 | 4 | 17 | 0.80 | 0.046 |
| **<u>Human-specific functional enhancers in hESC</u>** | 1619 | 10 | 14 | 0.86 | 0.034 |
| **Human accelerated DHS (haDHS)*** | 524 | 1 | 1 | 0.19 | 1 |
| **Human-specific CTCF binding sites in hESC*** | 575 | 2 | 3 | 0.52 | 0.410 |
| **Human-specific OCT4 binding sites in hESC*** | 2328 | 3 | 4 | 0.17 | 0.495 |
| **<u>Human-specific H3K4me3 peaks in prefrontal cortex</u>** | 406 | 5 | 5 | 1.23 | 0.027 |
| **<u>Human-specific NANOG binding sites in hESC</u>** | 816 | 14 | 19 | 2.33 | 5.04E-06 |
| **<u>hESC fixed human-specific regulatory regions (hESC-FHSRR)</u>** | 1932 | 14 | 29 | 1.50 | 0.00026 |
| **<u>Human-specific TE loci expressed in DLPFC</u>** | 4627 | 23 | 47 | 1.02 | 0.0046 |
| **<u>Human-specific gene expression in brain organoids</u>** | 1466 | 19 | 21 | 1.43 | 0.00087 |
| **<u>Radial glia</u>** | 947 | 17 | 19 | 2.01 | 3.43E-05 |
| **Excitatory neurons** | 517 | 5 | 5 | 0.97 | 0.0576 |
| **<u>Radial glia Down</u>** | 417 | 13 | 14 | 3.36 | 7.26E-07 |
| **Radial glia Up** | 531 | 4 | 5 | 0.94 | 0.062 |
| **Excitatory neurons Down** | 227 | 3 | 3 | 1.32 | 0.055 |
| **Excitatory neurons Up** | 291 | 2 | 2 | 0.69 | 0.261 |
| **Total number of genomic regions harboring HSRS** | 59732 | 248 | 442 | 0.74 | |

Legend: * the overlap of genomic coordinates of PtERV1 insertions and HSRS-harboring regions of this regulatory category is not statistically significant based on the hypergeometric distribution test; ** p values were estimated using the two-tailed Fisher's exact test compared to Human-specific STR contractions category; HSRS, human-specific regulatory sequences; WSSD, whole-genome shotgun sequence detection; underlined text denotes statistically significant categories estimated by both hypergeometric distribution test and two-tailed Fisher’s exact test;

**Table 2.** Evolutionary patterns of Chimpanzee, Gorilla, and Bonobo PtERV1 insertions intersecting genomic regions harboring HSRS.

| Taxa* | Number of PtERV1 loci | PtREV1 loci intersecting genomic regions harboring human-specific regulatory loci | Percent | P value** |
|---|---|---|---|---|
| **gorilla only** | 280 | 111 | 39.64 | 6.41E-20 |
| **chimp,gorilla** | 5 | 3 | 60.00 | 0.033 |
| **chimp,chimp,bonobo,gorilla,gorilla** | 1 | 1 | 100.00 | 0.168 |
| **chimp,chimp,bonobo,gorilla** | 5 | 0 | 0.00 | 0.399 |
| **chimp,chimp,bonobo** | 150 | 55 | 36.67 | 2.66E-09 |
| **chimp,bonobo** | 17 | 8 | 47.06 | 0.0029 |
| **chimp only** | 66 | 21 | 31.82 | 0.0012 |
| **bonobo,chimp,gorilla** | 1 | 0 | 0.00 | 0.832 |
| **bonobo,chimp** | 2 | 1 | 50.00 | 0.279 |
| **bonobo only** | 13 | 3 | 23.08 | 0.215 |
| **All PtERV1 insertions** | 540 | 203 | 37.59 | 3.24E-31 |
| **Gorilla all PtERV1 insertions** | 292 | 115 | 39.38 | 2.62E-20 |
| **Chimpanzee all PtERV1 insertions** | 247 | 89 | 36.03 | 1.71E-13 |
| **Bonobo all PtERV1 insertions** | 189 | 60 | 31.75 | 1.94E-07 |

Legend: * evolutionary patterns of 540 PtERV1 loci in genomes of non-human apes were reported by Kronenberg et al. (2018); ** p values were estimated using the hypergeometric distribution test considering the number of non-overlapping 10Kb regions in the human genome (308,829); the number of analyzed regions harboring HSRS (51,835); and corresponding numbers of the PtERV1 loci; no significant differences were observed between different categories;

**Table 2a.** Evolutionary patterns of Chimpanzee, Gorilla, and Bonobo PtERV1 insertions intersecting duplicated regions of human genome (hg38) defined by the whole-genome shotgun sequence detection (WSSD).

| Taxa* | Number of PtERV1 loci | PtREV1 loci intersecting genomic regions harboring human-specific regulatory loci | Percent | P value** |
|---|---|---|---|---|
| **gorilla only*** ** | 280 | 30 | 10.71 | 4.98E-11 |
| **chimp,gorilla** | 5 | 1 | 20.00 | 0.115 |
| **chimp,chimp,bonobo,gorilla,gorilla** | 1 | 1 | 100.00 | 0.026 |
| **chimp,chimp,bonobo,gorilla** | 5 | 0 | 0.00 | 0.879 |
| **chimp,chimp,bonobo** | 150 | 18 | 12.00 | 5.64E-08 |
| **chimp,bonobo** | 17 | 5 | 29.41 | 4.95E-05 |
| **chimp only*** ** | 66 | 16 | 24.24 | 7.73E-12 |
| **bonobo,chimp,gorilla** | 1 | 0 | 0.00 | 0.974 |
| **bonobo,chimp** | 2 | 0 | 0.00 | 0.950 |
| **bonobo only** | 13 | 0 | 0.00 | 0.714 |
| **All PtERV1 insertions** | 540 | 71 | 13.15 | 3.64E-29 |
| **Gorilla all PtERV1 insertions** | 292 | 32 | 10.96 | 6.50E-12 |
| **Chimpanzee all PtERV1 insertions** | 247 | 41 | 16.60 | 2.60E-21 |
| **Bonobo all PtERV1 insertions** | 189 | 24 | 12.70 | 1.26E-10 |

Legend: * evolutionary patterns of 540 PtERV1 loci in genomes of non-human apes were reported by Kronenberg et al. (2018); ** p values were estimated using the hypergeometric distribution test considering the number of non-overlapping 10Kb regions in the human genome (308,829); the number of analyzed duplication regions defined in human genome by WSSD (7,897); and corresponding numbers of the PtERV1 loci; no significant differences were observed between different categories; only PtERV1 loci intersecting duplication regions in hg38 are reported; *** indicated significantly different values for Gorilla only versus Chimpanzee only records (p = 0.0076; two-tailed Fisher's exact test);

**Table 3.** Mosaicism of evolutionary origins of genomic loci harboring various classes of human-specific mutations: Human-specific mutations target highly conserved sequences mapped to reference genomes of multiple species of non-human primates (95% sequence identity thresholds during both direct and reciprocal conversions).

| Classification category | Human genome | Chimpanzee | Percent | Bonobo | Percent | Gorilla | Percent | Orangutan | Percent | Gibbon | Percent | Rhesus | Percent | Mouse | Percent |
|---|---|---|---|---|---|---|---|---|---|---|---|---|---|---|---|
| **Fixed human-specific insertions*** | 11886 | 7138 | 60.05 | 6261 | 52.68 | 7104 | 59.77 | 3209 | 27.00 | 3268 | 27.49 | 2036 | 17.13 | 0 | 0.00 |
| **Fixed human-specific deletions** | 5891 | 4748 | 80.60 | 4487 | 76.17 | 4542 | 77.10 | 3569 | 60.58 | 3288 | 55.81 | 2923 | 49.62 | 96 | 1.63 |
| **Human-specific short tandem repeats (STR) expansions** | 4910 | 272 | 5.54 | 260 | 5.30 | 236 | 4.81 | 237 | 4.83 | 298 | 6.07 | 212 | 4.32 | 131 | 2.67 |
| **Human-specific short tandem repeats (STR) contractions** | 1464 | 1165 | 79.58 | 1115 | 76.16 | 1165 | 79.58 | 1031 | 70.42 | 927 | 63.32 | 867 | 59.22 | 85 | 5.81 |
| **Human-specific mutations associated with human-specific gene expression changes in excitatory neurons**** | 517 | 263 | 50.87 | 206 | 39.85 | 347 | 67.12 | 21 | 4.06 | 25 | 4.84 | 22 | 4.26 | 0 | 0.00 |
| **Human-specific mutations associated with human-specific gene expression changes in radial glia**** | 947 | 476 | 50.26 | 450 | 47.52 | 683 | 72.12 | 49 | 5.17 | 41 | 4.33 | 48 | 5.07 | 0 | 0.00 |
| **Duplicated regions in GRCh38 space defined by WSSD***** | 7897 | 5579 | 70.65 | 4969 | 62.92 | 5692 | 72.08 | 3895 | 49.32 | 3548 | 44.93 | 3236 | 40.98 | 597 | 7.56 |

Legend: *Results of the analyses of 10Kb regions centered at the exact sites of fixed human-specific insertions are reported; **Associated with changes in gene expression in human versus chimpanzee cerebral organoids; ***WSSD, whole-genome shotgun sequence detection;

**Table 4.** Mosaicism of evolutionary origins of genomic loci harboring candidate human-specific regulatory sequences associated with human-specific gene expression changes in human versus chimpanzee brain organoids. Only records remapped to the identical hg38 loci are reported.

| Classification category | Human genome | NHP genomes* | Percent | Chimpanzee | Percent | Bonobo | Percent | Gorilla | Percent | Orangutan | Percent | Gibbon | Percent | Rhesus | Percent | Mouse | Percent |
|---|---|---|---|---|---|---|---|---|---|---|---|---|---|---|---|---|---|
| **Human-specific mutations associated with human-specific gene expression changes in excitatory neurons**** | 517 | 283 | 54.74 | 183 | 64.66 | 143 | 50.53 | 249 | 87.99 | 9 | 3.18 | 16 | 5.65 | 14 | 4.95 | 0 | 0.00 |
| **Human-specific mutations associated with human-specific gene expression changes in radial glia**** | 947 | 526 | 55.54 | 326 | 61.98 | 282 | 53.61 | 472 | 89.73 | 21 | 3.99 | 24 | 4.56 | 27 | 5.13 | 0 | 0.00 |

Legend: *Mapped to highly conserved regions in genomes of non-human primates (NHP) identified based on 95% sequence identity thresholds during both direct and reciprocal conversions; **Associated with changes in gene expression in human versus chimpanzee cerebral organoids; percentage values for individual genomes of non-human species reflect fractions of all records scored in non-human primates;

**Table 5.** Mosaicism of evolutionary origins of candidate human-specific regulatory loci defined based on the mapping failure to both Chimpanzee and Bonobo reference genomes.*

| Classification category | Chimpanzee | Bonobo | Rhesus | Gibbon | Orangutan | Gorilla | Human |
|---|---|---|---|---|---|---|---|
| **Human-specific DLPFC-expressed transposons (n)**** | 0 | 0 | 1045 | 1148 | 1371 | 1688 | 4645 |
| **Percent** | 0.00 | 0.00 | 22.50 | 24.71 | 29.52 | 36.34 | 100.00 |
| **Human-specific hESC functional enhancers (n)** | 0 | 0 | 84 | 88 | 122 | 174 | 1619 |
| **Percent** | 0.00 | 0.00 | 5.19 | 5.44 | 7.54 | 10.75 | 100.00 |
| **Fixed human-specific regulatory regions (FHSRR) (n)** | 0 | 0 | 8 | 18 | 0 | 173 | 3273 |
| **Percent** | 0.00 | 0.00 | 0.24 | 0.55 | 0.00 | 5.29 | 100.00 |
| **hESC fixed human-specific regulatory regions (hESC-FHSRR)** | 0 | 0 | 2 | 2 | 0 | 70 | 1233 |
| **Percent** | 0.00 | 0.00 | 0.16 | 0.16 | 0.00 | 5.68 | 100.00 |
| **DHS fixed human-specific regulatory regions (DHS-FHSRR)** | 0 | 0 | 1 | 4 | 0 | 79 | 1104 |
| **Percent** | 0.00 | 0.00 | 0.09 | 0.36 | 0.00 | 7.16 | 100.00 |
| **Human-specific STR expansions (n)** | 0 | 0 | 35 | 47 | 30 | 16 | 1783 |
| **Percent** | 0.00 | 0.00 | 1.96 | 2.64 | 1.68 | 0.90 | 100.00 |
| **Fixed human-specific deletions (n)** | 0 | 0 | 4 | 6 | 6 | 9 | 58 |
| **Percent** | 0.00 | 0.00 | 6.90 | 10.34 | 10.34 | 15.52 | 100.00 |
| **Fixed human-specific insertions (n)** | 0 | 0 | 0 | 0 | 0 | 1 | 7 |
| **Percent** | 0.00 | 0.00 | 0.00 | 0.00 | 0.00 | 14.29 | 100.00 |
| **Human-specific NANOG binding sites (n)** | 0 | 0 | 1 | 4 | 0 | 52 | 540 |
| **Percent** | 0.00 | 0.00 | 0.19 | 0.74 | 0.00 | 9.63 | 100.00 |
| **Human-specific OCT4 binding sites (n)** | 0 | 0 | 0 | 5 | 0 | 19 | 1791 |
| **Percent** | 0.00 | 0.00 | 0.00 | 0.28 | 0.00 | 1.06 | 100.00 |
| **Human-specific CTCF binding sites (n)** | 0 | 0 | 0 | 0 | 0 | 14 | 478 |
| **Percent** | 0.00 | 0.00 | 0.00 | 0.00 | 0.00 | 2.93 | 100.00 |
| **Human-specific STR contractions (n)** | 0 | 0 | 7 | 9 | 11 | 15 | 199 |
| **Percent** | 0.00 | 0.00 | 3.52 | 4.52 | 5.53 | 7.54 | 100.00 |
| **All human-specific loci (n)** | 0 | 0 | 1187 | 1331 | 1540 | 2310 | 16730 |
| **Percent** | 0.00 | 0.00 | 7.10 | 7.96 | 9.21 | 13.81 | 100.00 |

Legend: *Mosaicism of evolutionary origins was defined based on the sequence conservation analyses of human-specific regulatory loci that failed conversions to both Chimpanzee and Bonobo genomes and manifested at least 95% sequence conservations during both direct and reciprocal conversions to genomes of other non-human primates (Gorilla; Orangutan; Gibbon; Rhesus); DLPFC, dorsolateral prefrontal cortex;

**Table 6.** Human-specific regulatory loci defined based on the conversion failure to reference genomes of Chimpanzee, Bonobo, Gorilla, Orangutan, Gibbon, and Rhesus.*

| Classification category | Human genome (hg38) | Mapping failures to both Chimpanzee and Bonobo genomes | Mapping failures to genomes of six non-human primates |
|---|---|---|---|
| **Human-specific hESC functional enhancers (n)** | 1619 | 1619 | 1343 |
| **Percent** | 100.00 | 100.00 | 82.95 |
| **Human-specific CTCF binding sites in hESC (n)** | 575 | 478 | 458 |
| **Percent** | 100.00 | 83.13 | 79.65 |
| **Human-specific OCT4 binding sites in hESC (n)** | 2328 | 1791 | 1731 |
| **Percent** | 100.00 | 76.93 | 74.36 |
| **Fixed human-specific regulatory regions (FHSRR) (n)** | 4249 | 3273 | 3035 |
| **Percent** | 100.00 | 77.03 | 71.43 |
| **hESC fixed human-specific regulatory regions (hESC-FHSRR)** | 1932 | 1233 | 1155 |
| **Percent** | 100.00 | 63.82 | 59.78 |
| **Human-specific NANOG binding sites (n)** | 816 | 540 | 483 |
| **Percent** | 100.00 | 66.18 | 59.19 |
| **Human-specific transposons expressed in dorsolateral prefrontal cortex (n)** | 4645 | 4612 | 2559 |
| **Percent** | 100.00 | 99.29 | 55.09 |
| **DHS fixed human-specific regulatory regions (DHS-FHSRR)** | 2116 | 1104 | 753 |
| **Percent** | 100.00 | 52.17 | 35.59 |
| **Human-specific STR expansions** | 4910 | 1783 | 692 |
| **Percent** | 100.00 | 36.31 | 14.09 |
| **Human-specific STR contractions** | 1464 | 199 | 183 |
| **Percent** | 100.00 | 13.59 | 12.50 |
| **Fixed human-specific deletions** | 5891 | 58 | 38 |
| **Percent** | 100.00 | 0.98 | 0.65 |
| **Fixed human-specific insertions** | 11886 | 7 | 4 |
| **Percent** | 100.00 | 0.06 | 0.03 |
| **Duplicated regions in GRCh38 space defined by WSSD** | 7897 | 75 | 52 |
| **Percent** | 100.00 | 0.95 | 0.66 |
| **All candidate human-specific regulatory loci (n)** | 50328 | 16697 | 12486 |
| **Percent** | 100.00 | 33.18 | 24.81 |

Legend: ** WSSD, whole-genome shotgun sequence detection; hESC, human embryonic stem cells; STR, short tandem repeats;

**Table 7.** Enrichment within human-specific regulatory networks of genes comprising expression signatures (GES) of human-specific neurodevelopmental transcriptional networks of excitatory neurons and radial glia.

**Networks of genes associated with expression of transposable elements (TE) in human dorsolateral prefrontal cortex**

| Classification category | Number of genes | Excitatory neurons | DOWN Humans vs Chimp | UP Humans vs Chimp | Radial glia | DOWN Humans vs Chimp | UP Humans vs Chimp |
|---|---|---|---|---|---|---|---|
| **Human genome** | 63677 | 384 | 165 | 219 | 668 | 285 | 383 |
| **Networks of genes associated with human DLPFC-expressed TE** | 22863 | 322 | 131 | 191 | 561 | 230 | 331 |
| **Percent** | 35.90 | 83.85 | 79.39 | 87.21 | 83.98 | 80.70 | 86.42 |
| **Enrichment**** | 1.00 | 2.34 | 2.21 | 2.43 | 2.34 | 2.25 | 2.41 |
| **P value*** | | 5.57E-84 | 2.61E-30 | 5.05E-56 | 3.08E-146 | 2.04E-54 | 9.93E-94 |

**GES of the Multi-lineage Markers Expressing (MLME) cells of human preimplantation embryo**

| Classification category | Number of genes | Excitatory neurons | DOWN Humans vs Chimp | UP Humans vs Chimp | Radial glia | DOWN Humans vs Chimp | UP Humans vs Chimp |
|---|---|---|---|---|---|---|---|
| **Human genome** | 63677 | 384 | 165 | 219 | 668 | 285 | 383 |
| **GES of the MLME cells of human embryo** | 12735 | 262 | 111 | 151 | 481 | 209 | 272 |
| **Percent** | 20.00 | 68.23 | 67.27 | 68.95 | 72.01 | 73.33 | 71.02 |
| **Enrichment**** | 1.00 | 3.41 | 3.36 | 3.45 | 3.60 | 3.67 | 3.55 |

| P value* | | 1.59E-93 | 1.50E-39 | 2.02E-55 | 3.95E-187 | 3.63E-84 | 1.22E-103 |
|---|---|---|---|---|---|---|---|
| **Regulatory networks of genes associated with human-specific structural variants*** ** | | | | | | | |
| **Classification category** | **Number of genes** | **Excitatory neurons** | **DOWN Humans vs Chimp** | **UP Humans vs Chimp** | **Radial glia** | **DOWN Humans vs Chimp** | **UP Humans vs Chimp** |
| **Human genome** | 63677 | 384 | 165 | 219 | 668 | 285 | 383 |
| **Genes associated with human-specific deletions and insertions** | 10992 | 123 | 53 | 70 | 125 | 59 | 66 |
| **Percent** | 17.26 | 32.03 | 32.12 | 31.96 | 18.71 | 20.70 | 17.23 |
| **Enrichment**** | 1.00 | 1.86 | 1.86 | 1.85 | 1.08 | 1.20 | 1.00 |
| **P value*** | | 6.75E-13 | 1.39E-06 | 4.66E-08 | 0.024 | 0.019 | 0.054 |

Legend: *, p values were estimate using the hypergeometric distribution test; **, expected values were estimated based on the number of genes in the human genome (63,677) and the number of genes in the corresponding category of human-specific regulatory networks; ***, this category of genes was reported in Kronenberg et al. (2018); TE, transposable genetic elements; hESC, human embryonic stem cells; DLPFC, dorsolateral prefrontal cortex; MLME, multi lineage markers expression;

**Table 8.** Enrichment within human-specific regulatory networks of genes comprising expression signatures (GES) of human-specific transcriptional networks of induced pluripotent stem cells (iPSC).

**Networks of genes associated with expression of transposable elements (TE) in human dorsolateral prefrontal cortex**

| Classification category | Number of genes | hiPSC network | DOWN hiPSC vs NHP iPSC | UP hiPSC vs NHP iPSC |
|---|---|---|---|---|
| **Human genome** | 63677 | 100 | 50 | 50 |
| **Networks of genes associated with human DLPFC-expressed TE** | 22863 | 76 | 40 | 36 |
| **Percent** | 35.90 | 76.00 | 80.00 | 72.00 |
| **Enrichment**** | 1.00 | 2.12 | 2.23 | 2.01 |
| **P value*** | | 2.72E-16 | 1.90E-10 | 1.77E-07 |

**GES of the Multi-lineage Markers Expressing (MLME) cells of human preimplantation embryo**

| Classification category | Number of genes | hiPSC network | DOWN hiPSC vs NHP iPSC | UP hiPSC vs NHP iPSC |
|---|---|---|---|---|
| **Human genome** | 63677 | 100 | 50 | 50 |
| **GES of the MLME cells of human preimplantation embryo** | 12735 | 52 | 22 | 30 |
| **Percent** | 20.00 | 52.00 | 44.00 | 60.00 |
| **Enrichment**** | 1.00 | 2.60 | 2.20 | 3.00 |
| **P value*** | | 8.91E-13 | 7.15E-05 | 5.72E-10 |

**Regulatory networks of genes associated with human-specific structural variants*****

| Classification category | Number of genes | hiPSC network | DOWN hiPSC vs NHP iPSC | UP hiPSC vs NHP iPSC |
|---|---|---|---|---|
| **Human genome** | 63677 | 100 | 50 | 50 |
| **Genes associated with human-specific deletions and insertions** | 10992 | 34 | 17 | 17 |
| **Percent** | 17.26 | 34.00 | 34.00 | 34.00 |
| **Enrichment**** | 1.00 | 1.97 | 1.97 | 1.97 |
| **P value*** | | 2.44E-05 | 2.03E-03 | 2.03E-03 |

Legend: Legend: *, p values were estimate using the hypergeometric distribution test; **, expected values were estimated based on the number of genes in the human genome (63,677) and the number of genes in the corresponding category of human-specific regulatory networks; ***, this category of genes was reported in Kronenberg et al. (2018); TE, transposable genetic elements; hESC, human embryonic stem cells; DLPFC, dorsolateral prefrontal cortex; MLME, multi-lineage markers expression; NHP, non-human primates; iPSC, induced pluripotent stem cells; hiPSC, human iPSC;

**Supplemental Table S1.** A catalog of human-specific genomic regulatory loci and networks reported to date.

| **Candidate human-specific regulatory loci** | **References** |
|---|---|
| Regions of human-specific loss of conserved regulatory DNA termed hCONDEL | McLean et al., 2011 |
| Human-specific epigenetic regulatory marks consisting of H3K4me3 histone methylation signatures at transcription start sites in prefrontal neurons | Shulha et al., 2012 |
| Human-specific transcriptional genetic networks in the frontal lobe | Konopka et al., 2012 |
| Conserved in humans regulatory DNA sequences designated human accelerated regions, HARs | Capra et al., 2013 |
| Fixed human-specific regulatory regions, FHSRR | Marnetto et al., 2014 |
| Human-specific transcription factor-binding sites, HSTFBS, and hESC enhancers | Glinsky, 2015-2018 |
| DNase I hypersensitive sites (DHSs) that are conserved in non-human primates but accelerated in the human lineage | Gittelman et al. 2015 |
| DNase I hypersensitive sites (DHSs) that are under accelerated evolution, ace-DHSs | Dong et al., 2016 |
| Fixed human-specific insertions | Kronenberg et al., 2018 |
| Fixed human-specific deletions | Kronenberg et al., 2018 |
| Human-specific short tandem repeats (STR) expansions | Kronenberg et al., 2018 |
| Human-specific short tandem repeats (STR) contractions | Kronenberg et al., 2018 |
| Human-specific mutations associated with human-specific gene expression changes in excitatory neurons | Kronenberg et al., 2018 |
| Human-specific mutations associated with human-specific gene expression changes in radial glia | Kronenberg et al., 2018 |
| Duplicated regions in GRCh38 space defined by WSSD | Kronenberg et al., 2018 |
| Human-specific hESC functional enhancers | Glinsky et al., 2018 |
| Human-specific gene expression signatures of induced pluripotent stem cells | Marchetto et al., 2013 |
| Human-specific segmental duplications | Dennis MY et al., 2017 |
| Genes with firmly established neurodevelopmental functions and well-documented genetic/genomic/epigenetic alterations of potential functional significance acquired within the human lineage after the divergence of humans and chimpanzees | Sousa, et al. 2017 |
| Chimp-biased developmental enhancers | Prescott et al., 2015 |
| Human-biased developmental enhancers | Prescott et al., 2015 |

**Supplemental Table S2.** Human-specific regulatory sequences (HSRS) analyzed in this contribution.

| Human-specific regulatory sequences (HSRS) | Number of loci | References |
|---|---|---|
| Human-specific STR contractions | 1464 | Kronenberg et al., 2018 |
| Human-specific STR expansions | 4910 | Kronenberg et al., 2018 |
| Fixed human-specific deletions | 5891 | Kronenberg et al., 2018 |
| Fixed human-specific insertions | 11886 | Kronenberg et al., 2018 |
| All regions of human-specific mutations | 24151 | Kronenberg et al., 2018 |
| Duplicated regions in GRCh38 space defined by WSSD | 7897 | Kronenberg et al., 2018 |
| Fixed human-specific regulatory regions (FHSRR) | 4249 | Marnetto et al., 2014 |
| Accelerated evolution-DHS (ace-DHS) | 3538 | Dong et al., 2016 |
| Human accelerated regions (HARs) | 2741 | Capra et al., 2013 |
| Chimp-biased developmental enhancers | 999 | Prescott et al., 2015 |
| Human-specific segmental duplications | 218 | Dennis et al., 2017 |
| Human-biased developmental enhancers | 996 | Prescott et al., 2015 |
| DHS Fixed human-specific regulatory regions (DHS-FHSRR) | 2116 | Marnetto et al., 2014 |
| Human-specific functional enhancers in hESC | 1619 | Glinsky et al., 2018; Glinsky, 2018; Barakat et al., 2018 |
| Human accelerated DHS (haDHS) | 524 | Gittelman et al. 2015 |
| Human-specific CTCF binding sites in hESC | 575 | Glinsky, 2015-2016; Kunarso et al., 2010 |
| Human-specific OCT4 binding sites in hESC | 2328 | Glinsky, 2015-2016; Kunarso et al., 2010 |
| Human-specific H3K4me3 peaks in prefrontal cortex | 406 | Shulha et al., 2012 |
| Human-specific NANOG binding sites in hESC | 816 | Glinsky, 2015-2016; Kunarso et al., 2010 |
| hESC fixed human-specific regulatory regions (hESC-FHSRR) | 1932 | Marnetto et al., 2014 |
| Human-specific TE loci expressed in DLPFC | 4627 | Guffani et al, 2018; Glinsky, 2018 |
| Total candidate HSRS | 59,732 | |

**Supplemental Table S3.** Human-specific regulatory sequences (HSRS) associated with human-specific gene expression changes in the cerebral organoid model of brain development.

| | |
|---|---|
| HSRS associated with human-specific gene expression in cerebral organoids | 1466 |
| HSRS associated with human-specific gene expression changes in Radial glia | 947 |
| HSRS associated with human-specific gene expression changes in Excitatory neurons | 517 |
| HSRS associated with genes Down-regulated in Radial glia of the human developing brain | 417 |
| HSRS associated with genes Up-regulated in Radial glia of the human developing brain | 531 |
| HSRS associated with genes Down-regulated in Excitatory neurons of the human developing brain | 227 |
| HSRS associated with genes Up-regulated in Excitatory neurons of the human developing brain | 291 |

Kronenberg et al., 2018. High-resolution comparative analysis of great ape genomes. Kronenberg et al., Science 360, 1085 (2018).

**Supplemental Table S4.** Mosaicism of evolutionary origins of 7,897 duplicated regions in GRCh38 space defined by whole-genome shotgun sequence detection (WSSD).

| Non-human species (NHS) | Number of highly conserved regions converted from/to hg38 | Percent | Loss of ancestral loci | Percent | Non-human species | Number of highly conserved regions converted from/to hg38 | Percent |
|---|---|---|---|---|---|---|---|
| Gorilla all records | 5571 | 81.61 | 1255 | 18.39 | Gorilla only records | 458 | 6.71 |
| Chimpanzee all records | 5484 | 80.34 | 1342 | 19.66 | Chimpanzee only records | 454 | 6.65 |
| Bonobo all records | 4862 | 71.23 | 1964 | 28.77 | Bonobo only records | 362 | 5.30 |
| Orangutan all records | 3744 | 54.85 | 3082 | 45.15 | Orangutan only records | 313 | 4.59 |
| Gibbon all records | 3331 | 48.80 | 3495 | 51.20 | Gibbon only records | 281 | 4.12 |
| Rhesus all records | 2951 | 43.23 | 3875 | 56.77 | Rhesus only records | 251 | 3.68 |
| Mouse all records | 451 | 6.61 | 6375 | 93.39 | Mouse only records | 58 | 0.85 |
| NHS all highly-conserved records | 6826 | 100.00 | 0 | 0.00 | NHS all species-specific highly-conserved records | 2177 | 31.89 |

Legend: reported only individual highly-conserved regions uniquely remapped to hg38 during reciprocal conversion from corresponding NHS genomes;

**Supplemental Table S5.** Mosaicism of evolutionary origins of genomic loci harboring candidate human-specific regulatory sequences associated with human-specific gene expression changes in human versus chimpanzee brain organoids.

| **Classification category** | Human genome | Mapped to highly conserved regions in genomes of non-human apes | Percent | Chimpanzee | Percent | Bonobo | Percent | Gorilla | Percent | Orangutan | Percent | Gibbon | Percent | Rhesus | Percent | Mouse |
|---|---|---|---|---|---|---|---|---|---|---|---|---|---|---|---|---|
| **Human-specific mutations associated with human-specific gene expression changes in excitatory neurons**** | 517 | 283 | 54.74 | **183** | 64.66 | **143** | 50.53 | **249** | 87.99 | **9** | 3.18 | 16 | 5.65 | **14** | 4.95 | 0 |
| **Human-specific mutations associated with human-specific gene expression changes in radial glia**** | 947 | 526 | 55.54 | **326** | 61.98 | **282** | 53.61 | **472** | 89.73 | **21** | 3.99 | 24 | 4.56 | **27** | 5.13 | 0 |

**Associated with changes in gene expression in human versus chimpanzee brain organoids; only records remapped to the identical loci in hg38 are reported;

Most significant fractions of candidate human-specific regulatory sequences associated with human-specific gene expression changes in both excitatory neurons (88.3%) and radial glia (89.7%) appear highly conserved in the Gorilla reference genome.

**Supplemental Table S6.** Conservation patterns of genomic regions harboring 11,866 fixed human-specific insertions.

| Region size/Species | Chimpanzee* | Percent | Bonobo | Percent | Gorilla | Percent | Orangutan | Percent | Rhesus | Percent | Mouse | Percent |
|---|---|---|---|---|---|---|---|---|---|---|---|---|
| 50 bp | 100 | 0.84 | 130 | 1.10 | 111 | 0.94 | 103 | 0.87 | 128 | 1.08 | 92 | 0.78 |
| 100 bp | 50 | 0.42 | 73 | 0.62 | 50 | 0.42 | 63 | 0.53 | 66 | 0.56 | 25 | 0.21 |
| 200 bp | 11 | 0.09 | 14 | 0.12 | 40 | 0.34 | 30 | 0.25 | 55 | 0.46 | 4 | 0.03 |
| 500 bp | 4 | 0.03 | 24 | 0.20 | 3 | 0.03 | 17 | 0.14 | 41 | 0.35 | 1 | 0.01 |
| 1100 bp | 53 | 0.45 | 63 | 0.53 | 49 | 0.41 | 28 | 0.24 | 52 | 0.44 | 0 | 0.00 |
| 2000 bp | 415 | 3.50 | 409 | 3.45 | 398 | 3.35 | 276 | 2.33 | 215 | 1.81 | 0 | 0.00 |
| 4000 bp | 1165 | 9.82 | 1057 | 8.91 | 1119 | 9.43 | 626 | 5.28 | 429 | 3.62 | 0 | 0.00 |
| 6000 bp | 1897 | 15.99 | 1665 | 14.03 | 1822 | 15.35 | 886 | 7.47 | 536 | 4.52 | 0 | 0.00 |
| 8000 bp | 6912 | 58.25 | 6036 | 50.87 | 6765 | 57.01 | 3061 | 25.80 | 1312 | 11.06 | 0 | 0.00 |
| 10000 bp | 7138 | 60.16 | 6261 | 52.76 | 7104 | 59.87 | 3209 | 27.04 | 2036 | 17.16 | 0 | 0.00 |

Legend: *Number of records that successfully completed direct and reciprocal conversions from/to hg38 and genomes of non-human primates and mouse using sequence identity threshold 95% of genomic regions centered at the insertion sites.